\documentclass[fleqn,usenatbib]{mnras}

\usepackage{graphics,rotating,multirow,bm,color,hyperref}
\usepackage{mathptmx}
\usepackage[T1]{fontenc}
\usepackage{ae,aecompl}
\usepackage{amsmath}
\usepackage{amssymb}
\usepackage{amsfonts,color,times,epstopdf,epsfig,url}
\usepackage[normalem]{ulem}

\usepackage{graphicx}	
\usepackage{amsmath}	
\usepackage{amssymb}	

\title[GGL in $f(R)$ gravity]{Galaxy-galaxy weak gravitational lensing in $f(R)$ gravity}

\author[]{
Baojiu Li$^1$\thanks{E-mail: baojiu.li@durham.ac.uk}, 
Masato Shirasaki$^2$\thanks{E-mail: masato.shirasaki@nao.ac.jp}
\\
$^{1}$Institute for Computational Cosmology, Department of Physics, Durham University, South Road, Durham DH1 3LE, UK\\
$^2$Division of Theoretical Astronomy, National Astronomical Observatory of Japan, 2-21-1 Osawa, Mitaka, Tokyo 181-8588, Japan
}

\date{Accepted XXX. Received XXX; in original form XXX}

\pubyear{2017}

\begin{document}
\label{firstpage}
\pagerange{\pageref{firstpage}--\pageref{lastpage}}
\maketitle

\begin{abstract}
We present an analysis of galaxy-galaxy weak gravitational lensing (GGL) in chameleon $f(R)$ gravity -- a leading candidate of non-standard gravity models. For the analysis we have created mock galaxy catalogues based on dark matter haloes from two sets of numerical simulations, using a halo occupation distribution (HOD) prescription which allows a redshift dependence of galaxy number density. To make a fairer comparison between the $f(R)$ and $\Lambda$CDM models, their HOD parameters are tuned so that the galaxy two-point correlation functions in real space (and therefore the projected two-point correlation functions) match. While the $f(R)$ model predicts an enhancement of the convergence power spectrum by up to $\sim30\%$ compared to the standard $\Lambda$CDM model with the same parameters, the maximum enhancement of GGL is only half as large and less than 5\% on separations above $\sim1$-$2h^{-1}$Mpc, because the latter is a cross correlation of shear (or matter, which is more strongly affected by modified gravity) and galaxy (which is weakly affected given the good match between galaxy auto correlations in the two models) fields. 
{We also study the possibility of reconstructing the matter power spectrum by combination of GGL and galaxy clustering in $f(R)$ gravity. We find that the galaxy-matter cross correlation coefficient remains at unity down to $\sim2$-$3h^{-1}$Mpc at relevant redshifts even in $f(R)$ gravity, indicating joint analysis of GGL and galaxy clustering can be a powerful probe of matter density fluctuations in chameleon gravity.}
The scale dependence of the model differences in their predictions of GGL can potentially allow to break the degeneracy between $f(R)$ gravity and other cosmological parameters such as $\Omega_m$ and $\sigma_8$. 
\end{abstract}




\section{Introduction}

One of the key, unanswered, questions in modern cosmology is the accelerated expansion rate of our Universe. Since its discovery almost 20 years ago \citep{riess98,perlmutter99}, it has been confirmed by various other observations, leading to the establishment of the concordance $\Lambda$ cold dark matter ($\Lambda$CDM) model, in which the Universe is dominated by a small cosmological constant $\Lambda$ which is solely responsible for the accelerated expansion at late times, while the formation of its structures has largely been shaped by dark matter component under the action of gravity, which is assumed to be described by Einstein's General Relativity (GR). While the simple $\Lambda$CDM model describes almost all cosmological observations very well \citep{WMAP9,Planck15}, currently there still lacks a satisfactory theoretical explanation for the smallness of $\Lambda$ required by observations, and this has led to significant effort in developing alternative scenarios, such as those which assume that GR is inaccurate and must be replaced by some modified gravity (MG) model on cosmological scales \citep[see, e.g.,][for recent reviews]{joyce15,Koyama2016}. Although a commonly accepted MG model still does not exist, and many of the MG models being proposed indeed introduce a $\Lambda$ through back door, studies in this field have so far led to various interesting possibilities of deviations from GR, which serve as useful testbeds of the validity of GR in cosmology. In recent and coming years, various imaging and spectroscopic galaxy surveys are producing high-quality data for a range of cosmological probes, with which we can hope to improve our understanding of the nature of the cosmic acceleration, along which it is hopeful to push the test of GR to much larger scales than previously attained.

Weak gravitational lensing \citep{bartelman01,refregier03,hoekstra08,kilbinger15} is one of the key cosmological probes that such galaxy surveys offer. It describes the effect that images of distant sources (e.g., galaxies or the CMB itself) are distorted by the intervening large-scale structures, which bend the paths of the photons emitted by the sources. Such bending is caused by visible plus dark matter, so that weak lensing offers a venue to detect the total matter distribution between the source and observer. Weak lensing can be observed in different ways, depending on the size of the lensing objects. At one extreme, the lensing effect by the largest bound objects in the Universe -- galaxy clusters -- is strong enough that it can be detected around individual clusters. At the other extreme, cosmic shear describes statistically the lensing effect of the entire matter distribution along the lines of sight. Other objects, such as cosmic voids and galaxies, produce lensing signals which are most often not strong or clean enough to allow a clear detection for individual lenses. However, with the stacking of a large number of lenses, a signal can be extracted and it can tell us how matter is distributed around such objects. In this paper we will focus on the lensing of background (source) galaxies by foreground (lens) galaxies, or galaxy-galaxy lensing 
{\citep[GGL;][]{1996ApJ...466..623B, 1998ApJ...503..531H, 2002MNRAS.335..311G,2004ApJ...606...67H, 2006MNRAS.368..715M}.}
For some recent works on testing cosmological models with cluster or void lensing, see \citet{cai15,cautun17,barreira15a,barreira15b,barreira17}. 

As a {\it gravitational} effect, weak {\it gravitational} lensing can naturally be used to test {\it gravitational} physics. On cosmological scales, a deviation from standard GR can leave detectable imprints on lensing observations in various ways. For example, it could change the expansion history through a different mechanism than $\Lambda$ to accelerate the expansion or via different best-fit cosmological parameters such as $\Omega_m$ and $H_0$, leading to different angular diameter distances to lenses and sources at given redshifts. It may have a different law of gravitational interaction, enhancing or reducing the clustering of matter at cosmological scales. It may also affect the propagation of photons, such as in the Galileon \citep{nicolis09,deffayet09}, nonlocal \citep{maggiore14,dirian14} and beyond Horndeski \citep{gleyzes15} gravity models. These effects, unfortunately, can have certain degeneracies between each other (and degenerate with the effects of cosmological parameters such as $\Omega_m$ and $\sigma_8$), and in this paper we focus on a simple model -- chameleon $f(R)$ gravity -- which practically does not modify the expansion history and photon propagation. This is one of the most popular classes of modified gravity models, which has the property of chameleon screening \citep{chameleon,chameleon2} to suppress the deviation from GR in regions of deep gravitational potential (such as our Solar System) and ensure that the theory passes local tests of gravity. 

The study of GGL in chameleon $f(R)$ gravity presented here is based on numerical simulations. Instead of doing a ray tracing to calculate the stacked lensing signal around galaxies, we will follow an equivalent approach by integrating the cross correlation function $\xi_{gm}(r)$ of galaxies and matter along the line of sight. The calculation is standard, but a significant effort of this study will be devoted to the construction of mock galaxy catalogues used to find $\xi_{gm}(r)$. The reason for this carefulness is twofold. Firstly, galaxies are observable and biased tracers of the underlying matter density field, and different populations of galaxies have different biases and density profiles around them. Therefore, it is important to know which galaxy population in observations should our simulation prediction of GGL signal be confronted to. Secondly (and which has perhaps not been emphasised enough so far), there is only one observable Universe, while there are many theories. For probes such as GGL, which involve the distributions of both source galaxies and the total matter field, the difference between the predictions of two models can come from differences in both. If, say, two models predict very different clustering of foreground galaxies, then one of them may already be incompatible with the spectroscopic observation used to identify these lenses, and the comparison of its GGL prediction to observations no longer makes sense. Of course, given the complex and poorly understood galaxy-mass and galaxy-halo connection, at this point it is premature to rule out the models studied here purely based on their predicted galaxy clustering. For example, commonly adopted frameworks to populate galaxies into simulations, such as halo occupancy distribution (HOD), abundance matching (SHAM) and semi-analytical modelling (SAM), usually have or can accept enough free parameters by tuning which the predicted and observed galaxy clustering can be matched (as we shall show below). Therefore, in this work we follow a pragmatic approach, by assuming that we have no idea whether GR or $f(R)$ gravity is the correct theory, and that for both of them the free parameters of the HOD (which we use to make galaxy mocks) can be tuned to produce acceptable clustering properties of the resulting galaxy catalogues, which will then be used to make predictions of GGL.

The layout of this paper is as follows. In \S\ref{sect:theory} we present a short description of the $f(R)$ gravity model studied here, a concise summary of weak gravitational lensing and GGL, and a detailed discussion of our numerical simulations and mock HOD galaxy catalogues. In \S\ref{sect:wl} we present the main results of this paper, including the power spectrum of cosmic shear, cross correlation between galaxies and matter, galaxy bias, and GGL in both GR and $f(R)$ gravity. We will also present a forecast of the potential to distinguish the two models using cosmic shear and GGL, for galaxy surveys such as Dark Energy Survey ({\sc des})\footnote{\url{https://www.darkenergysurvey.org}}, the Hyper Supreme Camera ({\sc hsc})\footnote{\url{http://hsc.mtk.nao.ac.jp/ssp/}} and Large Synoptic Survey Telescope ({\sc lsst})\footnote{\url{https://www.lsst.org}}. Finally, in \S\ref{sect:con} we sum up and discuss potential ways in which this work can be further improved.

Throughout this paper, we use the unit convention $c=1$ where $c$ is the speed of light. An overbar denotes the background value, and a subscript $_0$ denotes the present-day value, of a quantity. Greek indices $\mu, \nu, \cdots$ run over $0,1,2,3$ (space-time coordinates) while Latin indices $i,j,k,\cdots$ run over $1,2,3$ (space coordinates only). The Einstein summation convention is used across the paper unless otherwise stated. Also, we only consider a spatially flat universe.

\section{The theoretical framework}
\label{sect:theory}

We start with a concise description of $f(R)$ gravity and the model studied in this paper (\S\ref{subsect:MG_theory}), the formulae of weak gravitational lensing (\S\ref{subsect:wl}), and the $N$-body simulations used in the analyses and the catalogues of HOD galaxies which are needed to study their cross correlation with lensing (\S\ref{subsect:simulation}).

\subsection{The $f(R)$ gravity theory}
\label{subsect:MG_theory}

\subsubsection{Generic $f(R)$ gravity}

$f(R)$ gravity as an attempt to explain the accelerated late-time expansion of the Universe without invoking a cosmological constant was first proposed in \citet{carroll1,carroll2}. It is constructed by replacing the Ricci scalar $R$ in the standard Einstein-Hilbert action for GR with a function of $R$, $f(R)$:
\begin{equation}
S = \int{\rm d}^4 x \sqrt{-g}~\frac{1}{2}M_{\rm Pl}^2\left[R+f(R)\right],
\label{equ:E-H_action}
\end{equation}
in which $M_{\rm Pl}$ is the reduced Planck mass, with $M^{-2}_{\rm Pl}=8\pi G$ and $G$ being Newton's constant, $g$ is the determinant of the metric $g_{\mu\nu}$. The above is the gravitational action; the matter part is assumed to be the same as in standard $\Lambda$CDM, and so not explicitly given here.

A modified version of the Einstein equation can be derived by varying the action Eq.~(\ref{equ:E-H_action}) with respect to the metric $g_{\mu\nu}$: 
\begin{equation}
G_{\mu\nu}+f_{R}R_{\mu\nu}+\left[\square f_{R}-\frac{1}{2}f\right]g_{\mu\nu}-\nabla_\mu \nabla_\nu f_R=8\pi G T_{\mu\nu}.
\label{equ:E_equaiton}
\end{equation}
where $G_{\mu\nu}\equiv R_{\mu\nu}-\frac{1}{2}g_{\mu\nu}R$ is the Einstein tensor, $\nabla_\mu$ the covariant derivative, $f_R\equiv{\rm d}f/{\rm d}R$ is a (new) scalar degree of freedom in this theory, $\square\equiv\nabla^\alpha\nabla_\alpha$ the Laplancian and $T_{\mu\nu}$ the energy-momentum tensor for matter. 

Eq.~(\ref{equ:E_equaiton}) is a fourth-order equation in the metric tensor because the Ricii scalar $R$ itself contains second-order derivatives of the latter. It is convenient to cast it into a form involving the usual Einstein equation in GR plus the new scalar field (sometimes called the scalaron) $f_R$. The Klein-Gordon equation for $f_R$ can be derived as the trace of the modified Einstein equation:
\begin{equation}
\square f_R = \frac{1}{3}\left[R-f_R R+2f+8\pi G \rho_m\right],
\label{equ:fR}
\end{equation}
in which $\rho_m$ is the density for non-relativistic matter. We have neglected the contribution of relativistic matter species in this paper because we will focus on late times only.

The analysis in this study will be restricted to length scales much smaller than the horizon, in which case we can use the quasi-static approximation to drop all time derivatives of $f_R$ as compared with their spatial derivatives \citep[see, e.g.,][for a detailed discussion of this approximation]{bose15}. As a result of this simplification, Eq.~(\ref{equ:fR}) becomes
\begin{equation}
\vec\nabla^2 f_R = -\frac{1}{3}a^2\left[R(f_R)-\bar R+8\pi G(\rho_m-\bar \rho_m)\right],
\label{equ:fR_no_T}
\end{equation}
where $\vec\nabla$ is the spatial gradient and $a$ the scale factor ($a=1$ today). 

Also under the quasi-static and weak-field approximation, the Poisson equation that governs the Newtonian potential $\Phi$ is modified:
\begin{equation}
\vec\nabla^2 \Phi = \frac{16\pi G}{3}a^2(\rho_m-\bar \rho_m)+\frac{1}{6}\left[R(f_R)-\bar R\right].
\label{equ:X}
\end{equation}
The scalar field is what mediates a fifth force between massive particles in $f(R)$ gravity, and $f_R$ plays the role of its potential. To see this, we can combine Eqs.~(\ref{equ:fR_no_T}, \ref{equ:X}) to obtain the following equation:
\begin{equation}\label{eq:poisson}
\vec{\nabla}^2\Phi = 4\pi Ga^2(\rho_m-\bar \rho_m) + \frac{1}{2}\vec\nabla^2 f_R.
\end{equation}
Eq.~\eqref{eq:poisson} makes explicit the behaviour of the theory in the limit when $|f_R|\ll|\Phi|$: here we can to a good approximation neglect the effect of $f_R$ and recover the usual Poisson equation 
\begin{equation}
\vec{\nabla}^2\Phi = 4\pi G\delta\rho_m,
\end{equation}
in which $\delta\rho_m\equiv\rho_m-\bar{\rho}_m$ is the matter density perturbation. The GR solution $R=-8\pi G\rho_m$ is also recovered in this regime, where the fifth force is effectively suppressed, as the consequence of the so-called chameleon screening mechanism \citep{chameleon,chameleon2}.

As the opposite limit, when $|f_R|\geq|\Phi|$, we have $|\delta R|\ll\delta\rho_m$ with $\delta R\equiv R-\bar{R}$, and so the second term on the right-hand side of the modified Poisson equation, Eq.~(\ref{equ:X}), can be neglected:
\begin{equation}
\vec{\nabla}^2\Phi\approx\frac{16}{3}\pi G\delta\rho_m.
\end{equation} 
Compare this with Eq.~\eqref{eq:poisson}, we can notice a $1/3$ enhancement of the strength of gravit, independent of the functional form of $f(R)$. The exact form of $f(R)$, though, determines the transition between the two limiting regimes. We shall call these two regimes respectively the {\it screened} and {\it unscreened} regimes.

We can see that the chameleon mechanism works depending on the local Newtonian potential $\Phi$, and it efficiently screens the fifth force in environments where $\Phi$ is deep (i.e., $|\Phi|\gg|f_R|$). This can be qualitatively understood as follows: the fifth force is mediated by the scalar field $f_R$, which has a mass $m_{\rm s}$ given by
\begin{equation}
m^2_s = \frac{{\rm d}^2V_{\rm eff}(f_R)}{{\rm d}f_R^2},
\end{equation}
where the effective potential $V_{\rm eff}(f_R)$, due to the self-interaction of the scalar field and its interactions with (non-relativistic) matter, is given by
\begin{equation}
\frac{{\rm d}V_{\rm eff}(f_R)}{{\rm d}f_R} = \frac{1}{3}\left[R-f_R R+2f+8\pi G \rho_m\right].
\end{equation}
As a result of $m_s\neq0$ in general, the fifth force mediated by $f_R$ has the Yukawa form, with a potential $\sim r^{-1}\exp(-m_sr)$ where $r$ is the distance from a massive particle. The complicated form of $V_{\rm eff}(f_R)$ makes $m_s$ dependent on environment, and becomes heavy in deep $\Phi$, where the Yukawa force decays very quickly with distance, such that its effect is not felt beyond $\sim m_s^{-1}$. This is the origin of the chameleon screening, and this property can help the theory to pass stringent solar system tests, since it is expected that screening has effectively suppressed the fifth force to an undetected level at locations where such tests have been performed.


\subsubsection{The choice of $f(R)$ model}
\label{subsect:fR_theory}

The requirement of chameleon screening be in place for the model to pass solar system tests does not significantly constrain the possible functional form of $f(R)$. As a result, many different forms have been studied in the literature with differing details. While many of these are interesting, in practice it is both impossible and unnecessary to study all of them with equal detail. Instead, there are reasons why we should focus on a particular example which is to be studied in greater details.

First, there is currently not a fundamental theory to naturally motivate a functional form of $f(R)$ that leads to the cosmic acceleration, and therefore all choices of $f(R)$ used in the literature so far are phenomenological. There is no clear reason why any choice should be preferred over the others, apart from possibly an apparent simplicity in the functional form of $f(R)$. However, it is known that for general $f(R)$ models that realise an efficient chameleon screening, the background expansion history has to be very close to that of $\Lambda$CDM \citep[e.g.,][]{Brax2008,Wang2012,Ceron-Hurtado-2016}; in other words, a $\Lambda$ is usually introduced to the theory, possibly in an implicit way, regardless of the simplicity of $f(R)$. 

Second, as mentioned above, the different $f(R)$ models generally share some common features, such as chameleon screening in deep Newtonian potentials, and differ primarily in how efficient the screening is. However, given a functional form of $f(R)$, the screening efficiency also depends on the parameter used. At least qualitatively, the change of behaviours by varying the form of $f(R)$ can be mimicked by varying the parameters with a chosen $f(R)$. Instead of letting observations determine the form of $f(R)$, it is pragmatically more useful to use observations to determine to what extent deviations from GR as prescribed by $f(R)$ gravity are allowed. The latter task can be carried out by working on a particular model which is to be tested against as many observational data sets as possible, as precisely as possible.

For these reasons, this work is based on the model proposed by \citet[][hereafter HS]{hs2007}. This is the most well-studied example of $f(R)$ gravity, so that the results of this paper will be built upon various existing tests of this model. The particular functional form of $f(R)$ in this model makes it possible to implement an efficient algorithm to speed up simulations of it \citep{bose17}, which we adopt for the simulations used in this work. Additionally, this model is a representative example of classes of scalar-tensor-type modified gravity theories, in that by varying its parameters we can have a range of behaviours ranging from strong screening to no screening. 

The HS model has the following functional form of $f(R)$,
\begin{equation}\label{eq:hs}
f(R) = -\frac{c_1\left(-R/M^2\right)^n}{1+c_2\left(-R/M^2\right)^n}M^2,
\end{equation}
where $M$ is a parameter of mass dimension that is given by $M^2\equiv8\pi G\bar{\rho}_{m0}/3=H_0^2\Omega_m$, $H$ the Hubble rate, $\Omega_m$ the present-day density parameter for non-relativistic matter, and $c_1, c_2$ are dimensionless model parameters.

In the limit $|\bar{R}|\gg M^2$, $\bar{f}\equiv f(\bar{R})$ is approximately a constant $-\frac{c_1}{c_2}M^2$, so that $f_R$ and its derivatives are small. In this case, Eq.~(\ref{equ:fR}) can be simplified as 
\begin{equation}\label{equ:R_bg}
-\bar{R} \approx -2\bar{f}+8\pi G\bar{\rho}_m \approx 3M^2\left[\frac{2c_1}{3c_2}+a^{-3}\right].
\end{equation}
The background expansion rate is therefore close to that of $\Lambda$CDM, if we make mapping
$\frac{c_1}{c_2} = 6\frac{\Omega_\Lambda}{\Omega_m}$, where $\Omega_\Lambda\equiv1-\Omega_m$. For $\Omega_\Lambda\sim0.7$ and $\Omega_m\sim0.3$, we have $|\bar{R}|\sim40M^2\gg M^2$ today. As $|\bar{R}|$ increases with redshift, the approximation in Eq.~(\ref{equ:R_bg}) is good all the time. In this approximation, we have the following simplified relation between $f_R$ and $R$,
\begin{equation}\label{eq:temp}
f_R \approx -n\frac{c_1}{c_2^2}\left(\frac{M^2}{-R}\right)^{n+1}.
\end{equation}
This relation can be inverted to find $R(f_R)$, as the latter appears in field equations. Therefore, with choices of $n$ and $c_1/c_2^2$, as well as $\Omega_m, \Omega_\Lambda, H_0$, an $f(R)$ model can be fully specified. In the literature, instead of $c_1/c_2^2$, people usually use $f_{R0}$, the current value of $f_R$, as the model parameter. We have
\begin{equation}
\frac{c_1}{c_2^2} = -\frac{1}{n}f_{R0}\left[3\left(1+4\frac{\Omega_\Lambda}{\Omega_m}\right)\right]^{n+1}.
\end{equation}

In this paper, we shall focus on a particular choice of model parameters: $n=1$ and $f_{R0}=-10^{-5}$, which we refer to as F5. This choice of $f_{R0}$ is almost certainly incompatible with local gravity tests as the Milky Way galaxy is unlikely to be screened. However, the choice is not yet completely ruled out by cosmological observations, and for the GGL analysis we would like to choose a model that can maximise the difference from GR
\citep[see, e.g.,][for some recent studies on the current and potential constraints $f(R)$ gravity]{cai15,cataneo15,liu16,shirasaki16,shirasaki17,peirone17,cautun17}. 

\subsection{Weak gravitational lensing}
\label{subsect:wl}

Weak lensing is a matured field with a huge body of research works in the literature \citep[see, e.g.,][for some reviews]{bartelman01,hoekstra08,kilbinger15}. Here we only give a very quick catch-up of some essential equation to be used in the discussion below. For simplicity, we assume a flat Universe.

As photons pass through the large-scale structure in the Universe, their paths are bent by the latter, resulting in a change of their apparent angular position $\vec{\xi}_0\equiv\vec{\xi}(\chi=0)$ as seen by the observer at today (where $\chi$ is the comoving distance), as compared to the true one $\vec{\xi}_s\equiv\vec{\xi}(\chi_s)$, at the source (where $\chi_s$ is the comoving distance to the source). This is given by
\begin{equation}
\vec{\xi}_0 - \vec{\xi}_s = 2\int^{\chi_s}_0\frac{g\left(\chi,\chi_s\right)}{\chi^2}\vec{\nabla}_{\vec{\xi}}\Phi{\rm d}\chi, 
\end{equation}
where 
\begin{equation}
g\left(\chi,\chi_s\right) \equiv \frac{\chi}{\chi_s}\left(\chi_s-\chi\right),
\end{equation}
is the lensing kernel, and $\Phi\equiv\Phi\left(\chi,\vec{\xi}\right)$ is the lensing potential along the line of sight, and $\vec{\nabla}_{\vec{\xi}}$ is the two-dimensional derivative in the plane perpendicular to the l.o.s.. The resulting distortions of source images can be described by a $2\times2$ distortion matrix given by
\begin{eqnarray}
A_{ij} &=& \delta_{ij} - 2\int^{\chi_s}_0\frac{g\left(\chi,\chi_s\right)}{\chi^2}\nabla_{\xi_{0,i}}\nabla_{\xi_j}\Phi\left(\chi,\vec{\xi}\right){\rm d}\chi,~~~~~~~~~~~~~~~~~~~~~~\nonumber\\
&\approx& \delta_{ij} - 2\int^{\chi_s}_0\frac{g\left(\chi,\chi_s\right)}{\chi^2}\nabla_{\xi_{i}}\nabla_{\xi_j}\Phi\left(\chi,\vec{\xi}\right){\rm d}\chi,
\end{eqnarray}
where $i,j=1,2$, $\delta_{ij}$ is the Kronecker delta, $\nabla_{\xi_{0,i}}$ and $\nabla_{\xi_i}$ are respectively the derivative with respect to $\xi_i$ at $\chi=0$ and $\chi$. We have used $\nabla_{\xi_{0,i}}\approx\nabla_{\xi_i}$ in the second step of the above equation. {This is known as the Born approximation, and
its validity on weak lensing power spectrum and GGL has been studied in, e.g., \citet{hilbert08}}.

The lensing convergence $\kappa$, shear $(\gamma_1,\gamma_2)$ and rotation $\omega$ are related to the distortion matrix as
\begin{equation}
A = \begin{bmatrix}
    1-\kappa-\gamma_1 & -\gamma_2-\omega \\
    -\gamma_2+\omega & 1-\kappa+\gamma_1
\end{bmatrix},
\end{equation}
{where one can relate the convergence $\kappa$ to the density contrast $\delta\equiv\rho_m/\bar{\rho}_m-1$, by using the Poisson equation:}
\begin{equation}
\kappa = \frac{3}{2}\Omega_mH^2_0\int^{\chi_s}_0g\left(\chi,\chi_s\right)\nabla^2\frac{\delta}{a}{\rm d}\chi.
\end{equation}

Using the Limber approximation, the convergence power spectrum can be written as
\begin{equation}\label{eq:pk2cl}
C_{\kappa\kappa}(\ell) = \int^{\chi_s}_0{\rm d}\chi\frac{W(\chi)^2}{\chi^2}P_{\delta\delta}\left(k=\frac{\ell}{\chi},z(\chi)\right),
\end{equation}
where the lensing weight function $W(\chi)$ is given by
\begin{equation}
W(\chi)=\frac{3}{2}\Omega_mH_0^2g\left(\chi,\chi_s\right)\left[1+z(\chi)\right],
\end{equation}
$P_{\delta\delta}$ is the matter power spectrum, and $z(\chi)$ is the redshift corresponding to comoving distance $\chi$.

For GGL, we study the tangential shear profile around galaxies, which is given by
\begin{equation}
\gamma_t(r_p) = \frac{\Delta\Sigma(r_p)}{\Sigma_{\rm crit}}, \end{equation}
where $r_p$ is the projected distance from the galaxy, $\Delta\Sigma(r_p)\equiv\bar{\Sigma}(<r_p)-\Sigma(r_p)$ is the excess surface density defined as the difference between the mean surface (projected) mass density at $r<r_p$, $\bar{\Sigma}(<r_p)$, and the surface mass density at $r_p$, $\Sigma(r_p)$:
\begin{equation}
\bar{\Sigma}(<r_p) = \frac{1}{\pi r_p^2}\int_0^{r_p}\Sigma(r)2\pi r{\rm d}r,
\end{equation}
and $\Sigma_{\rm crit}$ is the critical surface mass density given as
\begin{equation}
\Sigma_{\rm crit} \equiv \frac{1}{4\pi G}\frac{\chi_s}{\chi_l\left(\chi_s-\chi_l\right)}\left[1+z(\chi_l)\right],
\end{equation}
in which $\chi_l$ is the comoving distance of the lens galaxy. Note that, given that the background expansion history is in practice indistinguishable in the GR and F5 models studied here, $\Sigma_{\rm crit}$ is the same for both models.

For individual galaxies, the lensing signal is weak, and so we consider the stacking of the tangential shear profile around many galaxies. In practice, the excess surface density profile can be computed as
\begin{eqnarray}\label{eq:excess}
\Delta\Sigma(r_p) &=& \rho_{\rm crit}\Omega_m\frac{2}{r_p^2}\int^{\infty}_{-\infty}{\rm d\chi}\int^{r_p}_0{\rm d}r\cdot r\xi_{gm}\left(\sqrt{r^2+\chi^2}\right)~~~~~~~~~~~~\nonumber\\
&& - \rho_{\rm crit}\Omega_m\int^{\infty}_{-\infty}{\rm d}\chi\xi_{gm}(r_p,\chi),
\end{eqnarray}
where $\xi_{gm}$ is the cross correlation function between galaxies and the matter density field.

\subsection{Simulations and halo/galaxy catalogues}
\label{subsect:simulation}

To accurately predict lensing effect in the nonlinear regime requires numerical simulations of the matter distribution in the Universe. To cross correlate this with galaxies requires mock galaxy catalogues with mimic the real galaxy distribution as observed by galaxy surveys. In this subsection, we shall describe in detail the simulations performed for this study, and halo and galaxy catalogues used in the analysis.

\subsubsection{$N$-body simulations}

The $f(R)$ gravity simulations used in our analysis have been run using the {\sc ecosmog} code \citep{ecosmog}, which is a modified version of the publicly available simulation code {\sc ramses} \citep{ramses}, by adding new modules to solve the $f(R)$ and Einstein equations. In this work we have used the optimised version of the code \citep{bose17}, which adopts a new algorithm to speed up simulations for HS $f(R)$ model with $n=1$ by a factor of up to $\sim20$. {\sc ramses}, and therefore {\sc ecosmog}, is efficiently parallelisated using {\sc mpi}, and is an example of the class of so-called adaptive-mesh-refinement (AMR) codes, which hierarchically refine a regular base mesh that covers the whole periodic simulation volume. In this way, it achieves the necessary high resolution in high density regions, without wasting substantial amount of computing resources in low density regions in which the demand for resolution is not as strong. The high resolution is also critical in order to resolve the fifth force effects in high density regions, such as in dark matter haloes, where the chameleon screening makes them weak (but not always negligble).

\begin{table}
\caption{The cosmological parameters for the models investigated in this work. $\Omega_m$ and $\Omega_\Lambda$ are respectively the present-day fractional density of matter and the cosmological constant (in the case of $f(R)$ gravity it is simply $1-\Omega_m$). $h=H_0/(100 {\rm km/s/Mpc})$ with $H_0$ being the present-day Hubble rate, $n_s$ is the spectral index of the primordial density fluctuations, $A_s$ the amplitude of the primordial power spectrum, and $\sigma_8$ the root-mean-squared (RMS) linear density fluctuation in spheres of radius $8h^{-1}$Mpc at $z=0$ (the value quoted here is for $\Lambda$CDM model only, as F5 has a different value). 
$\Omega_b$ is the baryon density used for the linear matter power spectrum, to generate the initial conditions only.
}
\begin{tabular}{@{}lll}
\hline\hline
parameter & physical meaning & value \\
\hline
$\Omega_m$  & present fractional matter density & $0.2819$ \\
$\Omega_b$  & present fractional baryon density & $0.0461$ \\
$\Omega_{\Lambda}$ & $1-\Omega_m$ & $0.7181$ \\
$h$ & $H_0/(100$~km~s$^{-1}$Mpc$^{-1})$ & $0.6970$ \\
$n_s$ & primordial power spectral index & $0.9710$ \\
$\log_{10}A_{s}$ & amplitude of the primordial power spectrum & $-8.622$ \\ 
$\sigma_{8}$ & RMS density fluctuation at $8h^{-1}$Mpc for $\Lambda$CDM & $0.8178$ \\
\hline
$n$ & HS $f(R)$ parameter & $1.0$ \\
$f_{R0}$ & HS $f(R)$ parameter & $-10^{-5}$ \\
\hline
\hline
\end{tabular}
\label{table:cos_params}
\end{table}

\begin{table}
\caption{The simulation specifications. $L_{\rm box}$ is the  simulation box size, $N_p$ is the number of simulation particles, $m_{p}$ the simulation particle mass, $N_{\rm r}$ the number of realisations for each box size and $N_{\rm s}$ the number of particle outputs (which are spaced every $75h^{-1}$Mpc in comoving distance from today). $z_{\rm ini}$ is the starting redshift of the simulations, and the initial condition is generated using the {\sc 2lptic} code. Out of the $N_s$ snapshots, 33 are between $z=[0,1]$.} 
\begin{tabular}{@{}lllllll}
\hline\hline
$L_{\rm box}$ [$h^{-1}$Mpc] & $z_{\rm ini}$ & $N_p$ & $m_p$ [$h^{-1}M_\odot$] & $N_{\rm r}$ & $N_{\rm s}$ & IC \\
\hline
$450$ & $49.0$ & $1024^3$ & $6.64\times10^9$ & $1$ & $37$ & 2LPT \\
$900$ & $36.0$ &$1024^3$ & $5.31\times10^{10}$ & $1$ & $37$ & 2LPT \\
\hline
\hline
\end{tabular}
\label{table:sim_params}
\end{table}

The cosmological parameters, listed in Table~\ref{table:cos_params}, have been chosen from the best-fit \citep{WMAP9} WMAP9 $\Lambda$CDM model, and their physical meanings are explained in the table caption. The technical parameters for the simulations are given in Table \ref{table:sim_params}. All simulations have $1024^3$ particles. 

The initial conditions of the simulations are generated using the {\sc 2lptic} code which is based on second-order Lagrangian perturbation theory \citep{crocce06}. Since we keep the simulation particle number $N_p$ fixed to be $1024^3$, for our larger boxes the mass resolution is relatively low, and so following \citet{shirasaki15} we compensate this by starting those simulations at relatively low redshifts. For each F5 simulation, we run a $\Lambda$CDM simulation with exactly the same cosmological parameters and simulation specifications for comparison; the $\Lambda$CDM simulations also start from the same initial conditions as their F5 counterparts, because at $z_{\rm ini}\gg1$ the difference between the two models is negligible. We only have one realisation of simulation for each box size.

\subsubsection{Halo and galaxy catalogues}

The dark matter haloes used in this paper are found using the publicly available phase-space friend-of-friend halo finder {\sc rockstar} \citep{rockstar}, and we have chosen $M_{200c}$ as the mass definition, in which the subscript means that the average mass density within halo radius $R_{200c}$ is 200 times the critical density $\rho_c$.

We populate dark matter haloes using halo occupation distribution \citep[HOD;][]{hod,zheng2005}. In the simplest form of this model, the mean number density of galaxies in a host dark matter halo is a function of the halo mass $M$:
\begin{equation}\label{eq:Ng}
\langle N_{g}|M\rangle = \langle N_{c}|M\rangle\left[1+\langle N_{s}|M\rangle\right],
\end{equation}
where $N_g$, $N_c$ and $N_s$ are respectively the number of all, central and satellite galaxies, given as
\begin{equation}\label{eq:Nc}
\langle N_c|M\rangle = \frac{1}{2}\left[1+{\rm erf}\left(\frac{\log M-\log M_{\rm min}}{\sigma_{\log M}}\right)\right],
\end{equation}
\begin{equation}\label{eq:Ns}
\langle N_s|M\rangle = \left(\frac{M-M_0}{M_1}\right)^\alpha\Theta\left(M-M_{0}\right),
\end{equation}
where ${\rm erf}(x)$ and $\Theta(x)$ are respectively the error and Heaverside step functions. The model has five parameters: $M_{\rm min}, M_0, M_1$, $\sigma_{\log M}$ and $\alpha$, which give it great freedom to tune the  galaxy catalogues to match their statistical properties to observables. 

To implement the HOD model, we sift through the halo catalogue where subhaloes are eliminated, and for each halo let it host a central galaxy if $u\leq\langle N_c|M\rangle$ where $u$ is a random number generated from a uniform distribution between $[0,1]$. The number of satellite galaxies is set to a random number generated from the Poisson distribution with mean $\langle N_s\rangle$, and the satellite galaxies are radially distributed within the dark matter halo following the Navarro-Frenk-White \citep[NFW;][]{NFW} profile, using the concentration parameters measured by {\sc rockstar}. In the rare cases where a halo has no central galaxy but does have satellites, we promote the first satellite galaxy to a central galaxy.

Although this paper does not aim to compare theoretical predictions with real observations, we still want the HOD catalogues to bear a certain degree of reality. To this end, we focus on galaxies at relatively low redshift (where the model difference is expected to be larger) and make the resulting galaxies satisfy a redshift distribution similar to that of of the Low Redshift Sample (LOWZ) of BOSS data release 11. Instead of down-sampling generated HOD catalogues, we follow \cite{manera2015} to allow a redshift dependence of the HOD mass parameters as follows:
\begin{equation}\label{eq:M_evol1}
\log M_{\rm min} = \log M^\ast_{\rm min} + Sn(z)/\left(10^{-4}\right),
\end{equation}
\begin{equation}
\label{eq:M_evol2}\log M_{0,1} = \log M^\ast_{0,1} + Tn(z)/\left(10^{-4}\right),
\end{equation}
in which $n(z)$ is the target galaxy number density for redshift $z$, $S=-0.925$, $T=-0.928$, and $M^\ast$ are the respective mass parameter values at $n^\ast=2.98\times10^{-4}$. Note that $M_1/M_0$ has no time evolution. We consider galaxies 
in the redshift range of $0.16\lesssim z\lesssim 0.43$. 
We use 29 redshift bins in this range, with equal comvoing thickness, with 3 bins being taken from each of the 10 snapshots of this box. For the 3 bins from the same simulation snapshot, the evolution of $n(z)$ is only down to Eqs.~(\ref{eq:M_evol1}, \ref{eq:M_evol2}), while for the bins from different snapshots $n(z)$ is also affected by the fact that different halo catalogues have been used to build the HOD catalogues.

Should the HOD catalogues in F5 have been constructed with the same HOD parameters as in their $\Lambda$CDM counterparts, there would generally be a difference of $\sim10$-$20$\% in the number density and clustering of galaxies. Because there is only one observed galaxy catalogue in the Universe (assuming an ideal full-sky survey), if we do not know which is the correct cosmological model, we can only demand that both models make predictions that agree with observations. For this reason, we have opted to tune the HOD parameters for F5 so that the resulting galaxy catalogues match the number density and clustering of their $\Lambda$CDM counterparts. The assumption that F5 and GR should have different HOD parameters is not unreasonable, given their different dark matter evolutions and galaxy assembly histories. This choice indeed also helps to fix the galaxy clustering and single out the expected difference of the galaxy-galaxy lensing signals in these two models.

The tuning of F5 HOD parameters has been performed using a search with the Nelder-Mead simplex algorithm through the 5-dimensional parameter space. The 3D real-space correlation functions in the F5 and GR HOD catalogues are measured between comoving separations of $0.6$ and $60h^{-1}$Mpc, in 40 equally-spaced logarithmic bins. The r.m.s.~difference between the two models is calculated with equal weight 1 for all bins. To try to make the two models have similar $n(z)$, the relative difference in their $n(z)$ values is also added into the r.m.s., with a weight 8. The code then searches through the 5D parameter space looking for the smallest r.m.s.~difference ($\chi^2$). The search stops when $\chi^2<0.03$, meaning that the overall agreement (as defined in the above way) is better than 3\% (for some redshift bins better than 2\%). We have not attempted to do better than this accuracy, or perform a full parameter search using Markov chain Monte Carlo, because it is not the purpose of this paper to study in great detail the HOD model in the context of modified gravity, and because it is in general difficult to do better than this level of accuracy (for example, the galaxy clustering may change at few percent level if a different set of random seeds is used in generating the HOD catalogues). For the same reason, we shall not list the HOD parameters in this paper.

\begin{figure*}
\includegraphics[width=18cm]{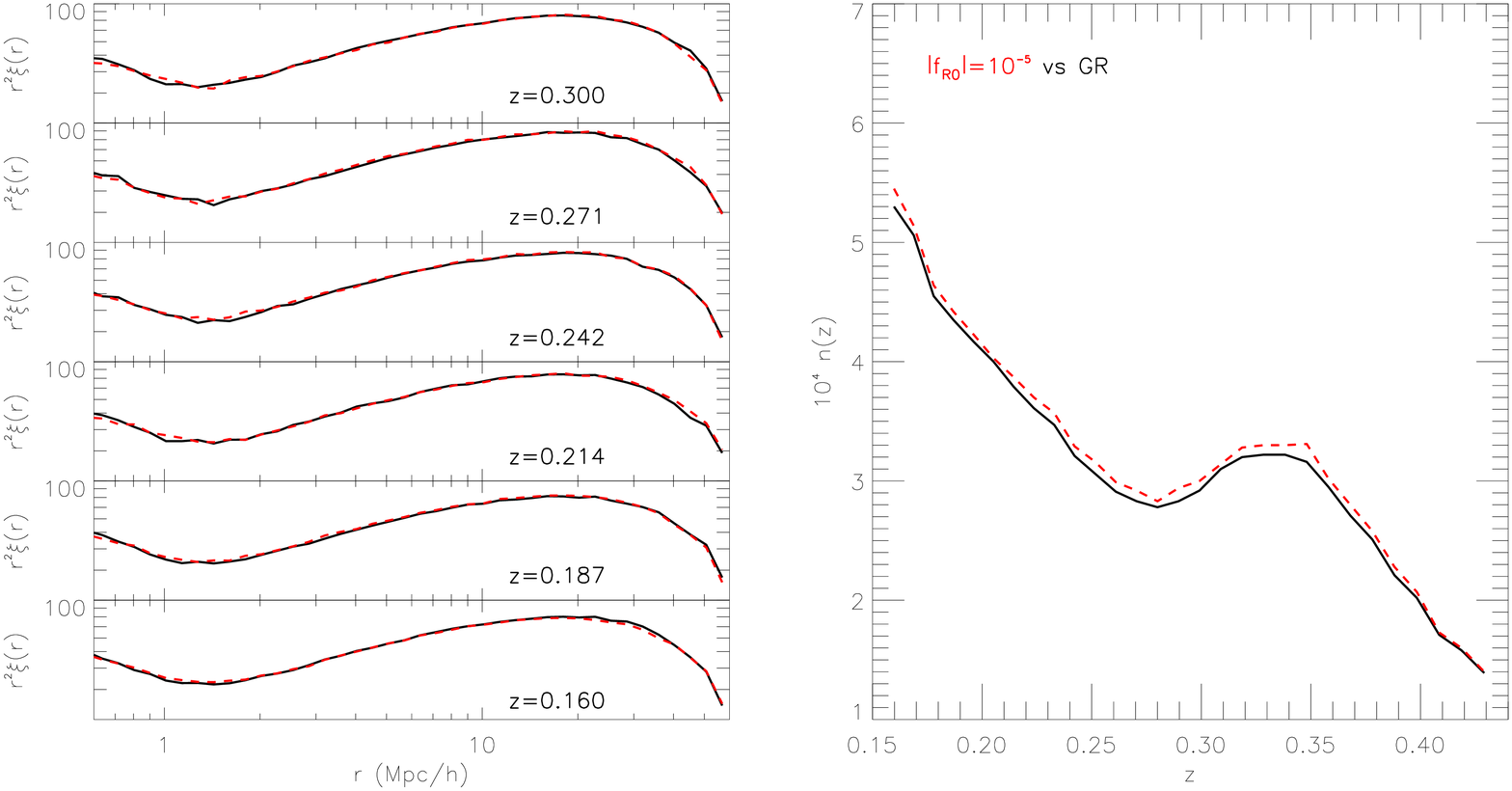}
\caption{(Colour Online) {\it Left Panels}: Comparison of the GR (black solid) and F5 (red dashed) correlation functions, after tuning the HOD parameters for F5 such that the two models predict similar galaxy number densities and clusterings. For clearness we only show results for 6 of the 29 redshift bins from the $900h^{-1}$Mpc box. Each bin is taken from HOD catalogues built using a different output snapshot. {\it Right Panel}: Comparison of the predicted GR (black solid) and F5 (red dashed) galaxy redshift distributions. See the text for more details on how the galaxy numbers in each redshift bin are determined for GR and F5.}
\label{fig:tpcf}
\end{figure*}

In the left panel of Figure \ref{fig:tpcf} we compare the measured 3D real-space correlation functions $\xi(r)$ as a function of galaxy separation $r$, respectively for F5 (red dashed lines) and GR (blue solid) at 6 of the 29 redshift bins. In the right panel of Fig~\ref{fig:tpcf} (the same colour scheme) shows the redshift distribution of the HOD galaxies for the two models -- again there is a good agreement (generally better than $4$\%) after tuning the HOD parameters for F5.

\section{Weak gravitational lensing in f(R) gravity}
\label{sect:wl}

In this section we present the main result of this research, beginning with an analysis of the weak lensing convergence. The convergence power spectrum in $f(R)$ gravity, amongst various other things, have been studied in a few works \citep{shirasaki15,shirasaki17,higuchi16,tessore15,Pratten16}, using various techniques. We revisit this topic in \S\ref{subsect:convergence} with a detailed decomposition of the convergence power spectrum $C_{\kappa\kappa}(\ell)$ in terms of the length scales and times where the contributions to the relative difference between $f(R)$ gravity and GR come from. The analysis not only serves as a sanity check of the simulations and their corresponding particle snapshots, but also will be used for comparison with results on galaxy-galaxy lensing from \S\ref{subsect:ggl}, as they are based on the same set of simulations.

\begin{figure*}
\includegraphics[width=18cm]{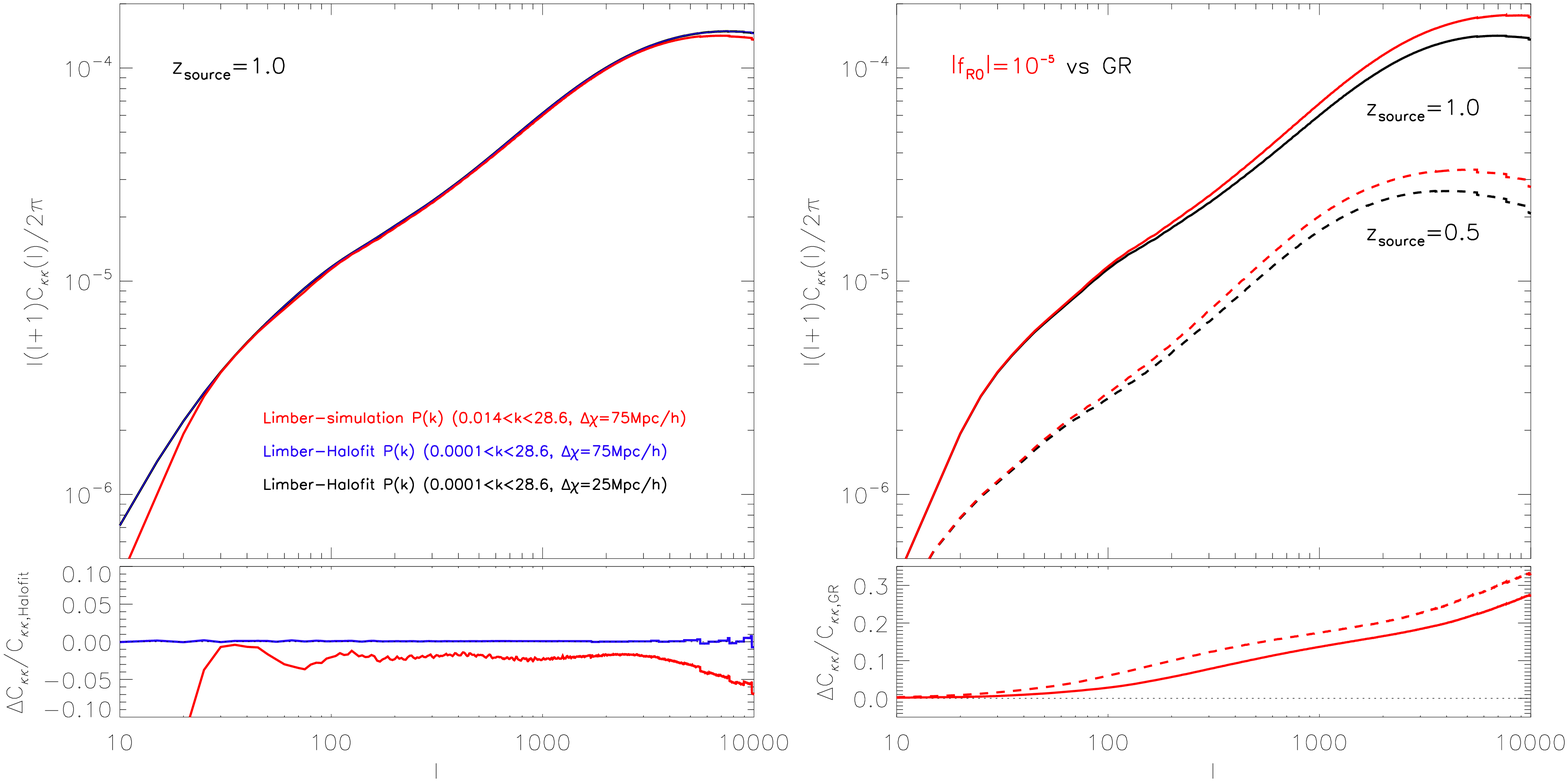}
\caption{(Colour Online) {\it Left Panels}: the effect of choosing a coarser -- $\Delta\chi=75h^{-1}$Mpc in blue solid vs $\Delta\chi=25h^{-1}$Mpc in black solid -- sampling of $P(k,z)$ (generated using {\sc halofit}) when using the Limber approximation Eq.~(\ref{eq:pk2cl}) to compute the convergence power spectrum $C_{\kappa\kappa}(\ell)$. The solid red line shows the result of doing the same integration with $\Delta\chi=75h^{-1}$Mpc, but for the simulated $P(k,z)$. The lower sub-panel shows the relative differences of the other two cases w.r.t.~the integration of {\sc halofit} $P(k,z)$ with $\Delta\chi=25h^{-1}$Mpc. {\it Right Panels}: the convergence power spectra for F5 (red) and GR (black), using two source redshifts $z_{\rm source}=1.0$ (solid) and $0.5$ (dashed) -- all results are from integrating the simulated $P(k,z)$ following Eq.~(\ref{eq:pk2cl}). The lower sub-panel shows the relative differences between F5 and GR. See the main text for discussions of this plot.}
\label{fig:sim_halofit}
\end{figure*}

\subsection{The convergence field}
\label{subsect:convergence}

The weak lensing convergence is an important physical quantity for weak lensing research: it can be calculated directly from a simulation by a projection of the density field, and in the regime of weak lensing its power spectrum coincides with that of the cosmic shear, which is a directly-observable quantity. It has been one of the most widely used quantities in theoretical analyses (see reference above for examples in $f(R)$ gravity).

\subsubsection{The convergence power spectrum in $f(R)$ gravity}

In this work we compute the weak lensing convergence power spectrum $C_{\kappa\kappa}(\ell)$ by directly integrating the 3D matter power spectrum $P_{\delta\delta}(k,z)$ measured from the simulations, following the Limber approximation Eq.~(\ref{eq:pk2cl}), instead of using full ray tracing. This is partly because the former approach essentially samples all the relevant $k$-modes provided by the simulation, while ray tracing only samples those $k$ modes that fall into the designed lightcone. While the latter more closely mimics the real Universe, the former gives smoother theoretical curves. Eq.~(\ref{eq:pk2cl}), together with an accurate prediction of the nonlinear $P_{\delta\delta}$ such as from {\sc halofit} \citep{halofit,halofit2}, is a powerful tool to predict $C_{\kappa\kappa}$.

The nonlinear $P_{\delta\delta}$ used in this work are measured using the publicly available {\sc powmes} code \citep{powmes} with a Fast Fourier Transform (FFT) grid of size $2048^3$. As mentioned above, our simulations have outputs at 33 snapshots between $z=1$ and $z=0$ equally spaced in comoving distance ($\Delta\chi=75h^{-1}$Mpc), and therefore this is the finest grid in $z$ (or in $\chi$) that our integration of Eq.~(\ref{eq:pk2cl}) can be carried out. We used the 5-point Newton-Cotes formula for the numerical integration over $\chi$, and values of $P_{\delta\delta}(k)$ at a given redshift $z(\chi)$ were obtained by linear interpolation of the corresponding values at $k_i, k_{i+1}$ with $k_{i}<k<k_{i+1}$, where $k_i$ is the $i$-th grid point in the {\sc powmes} output. To check that such a coarse spacing in $\chi$ is not causing substantial numerical errors, we made a test using a finer sampling of $P_{\delta\delta}(k,z(\chi))$ in the $\chi$ direction, with a comoving thickness of $25h^{-1}$Mpc, generated using {\sc halofit}. We then integrated this sample (1) using all $\chi$ bins, and (2) using one from every three neighbouring bins (the latter mimicking the simulation binning of $\Delta\chi=75h^{-1}$Mpc). The results of these two tests are shown as black and blue solid lines in the upper left panel of Figure \ref{fig:sim_halofit}, and their relative difference is presented as the blue solid line in the lower left panel. The tests showed that the difference is much smaller than $1\%$, and that $\Delta\chi=75h^{-1}$Mpc is fine enough for our study. We have also checked that the convergence power spectrum (for GR) computed in this way agreed well with the result of {\sc camb} \citep{camb} for the same cosmology, though note that to match the $k$ range of the simulated $P_{\delta\delta}(k,z)$ we have limited the maximum $k$ in the integration to $k_{\rm max}=28.6~h$Mpc$^{-1}$, which affects $C_{\kappa\kappa}$ a little bit at $\ell\sim10^4$.

The red solid line in the upper left panel of Fig.~\ref{fig:sim_halofit} is the result of the same integration of the simulated $P_{\delta\delta}(k,z)$ for GR, and the red solid line in the lower left panel is the relative difference from the integration of {\sc halofit} $P_{\delta\delta}(k,z)$. The simulation result peels off at $\ell\lesssim20$ because the lowest $k$ values used in the integration is $k_{\rm min}\sim0.014~h$Mpc$^{-1}$, and the same feature appeared in the {\sc halofit} results if the same $k_{\rm min}$ was applied there. We notice that the simulation and {\sc halofit} results agree reasonably well, with the former lower by $\sim2\%$ in a wide range of $\ell$, possibly due to sampling variance (recall that the simulation box is $450h^{-1}$Mpc). The difference between the two further changes to $5-6\%$ when $\ell$ approaches $10^4$, due to slight loss of resolution by the simulation.

The right panels of Figure~\ref{fig:sim_halofit} compare the convergence power spectra of F5 (red) and GR (black), for two lensing source redshifts, $z_{\rm source}=1.0$ (solid) and $z_{\rm source}=0.5$ (dashed). For both source redshifts, an enhancement of order $10\%-30\%$ between $\ell\sim100$ and $\ell\sim10^4$ is found, which increases with $\ell$, and the relative difference is larger in the case of $z_{\rm source}=0.5$.

\subsubsection{The connection between matter and convergence spectra}

\begin{figure*}
\includegraphics[width=18.5cm]{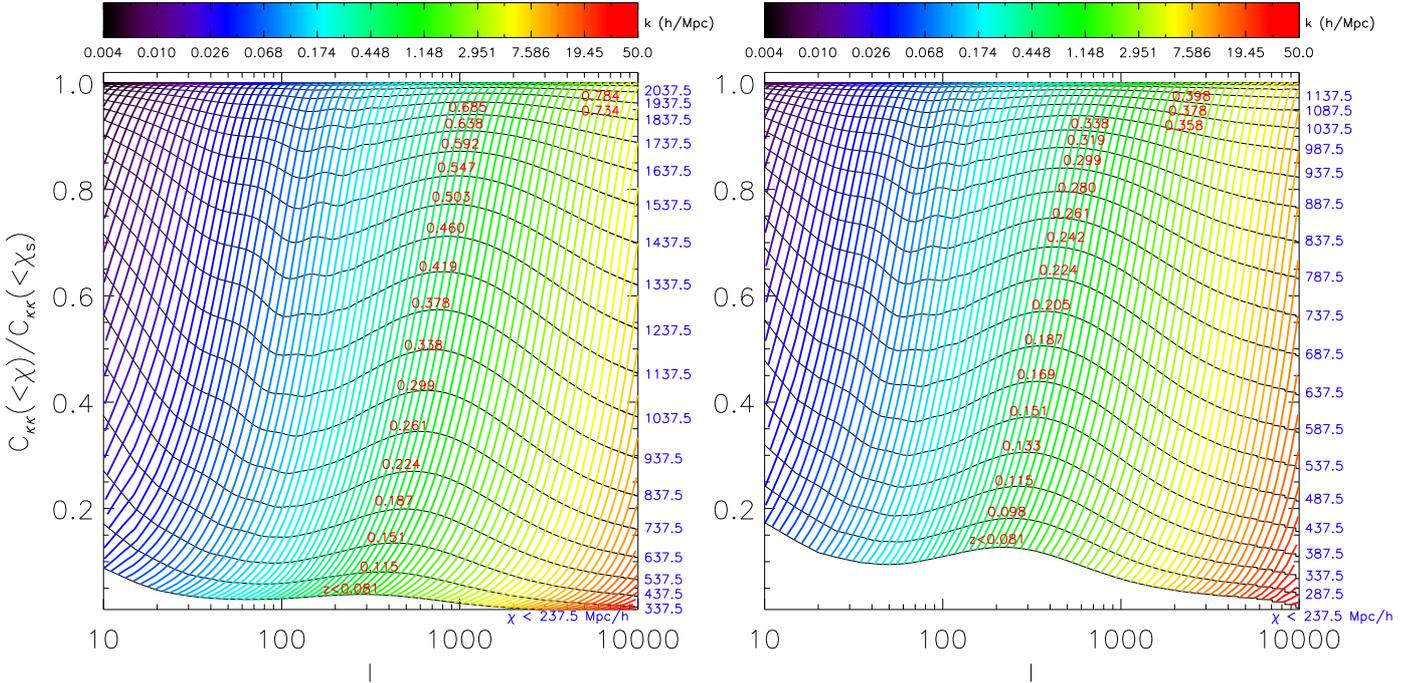}
\caption{(Colour Online) A detailed decomposition showing how much the different time (redshift) intervals and $k$-modes in the nonlinear matter power spectrum $P_{\delta\delta}(k,z)$ contribute to the convergence power spectrum $C_{\kappa\kappa}(\ell)$ at a given $\ell$. The decomposition is done for two source redshifts: $z_{\rm source}=1.0$ (left panel) and $0.5$ (right panel). The vertical axis is the fractional contribution from distances smaller than a given $\chi$, defined as $C_{\kappa\kappa}(<\chi)/C_{\kappa\kappa}(<\chi_{s})$, and the black solid lines are the values for a selection of $\chi$ values (indicated in blue; the corresponding redshift values are indicated in red) in step of $100h^{-1}$Mpc (for $z_{\rm source}=1$) and $50h^{-1}$Mpc (for $z_{\rm source}=0.5$). The colour-coded solid lines (roughly) in the vertical direction are curves with constant $k$ values that contribute to a given $\ell$ at a given time $z$ (or $\chi$), with $k=\ell/r(\chi)$. The value of $k$ for each of these curves can be found using the colour-bar on the top, which runs from black ($k=0.004~h$~Mpc$^{-1}$) to red ($k=50~h$~Mpc$^{-1}$). Note that each $\ell$ in $C_{\kappa\kappa}(\ell)$ receives contributions from a limited range of $k$ (which shifts to larger values for decreasing $z_{\rm source}$), and that most contributions are from low and middle redshifts.}
\label{fig:decomp}
\end{figure*}

To gain a clearer understanding of the behaviour shown in the right panels of Figure \ref{fig:sim_halofit}, recall from Eq.~(\ref{eq:pk2cl}) that $C_{\kappa\kappa}(\ell)$ at a given angular scale $\ell$ gets contribution from various $k$ modes of the nonlinear $P_{\delta\delta}$ from different times. It is therefore useful to decompose the contributions from the different times and $k$ modes, onto which we can readily map the relative differences in $P_{\delta\delta}(k,z)$ themselves. Figure \ref{fig:decomp} shows the results from such an attempt for $z_{\rm source}=1$ (left panel) and $z_{\rm source}=0.5$ (right panel). In both cases, the plot is the fractional contribution (black solid curves) from distances below a specific $\chi$ (indicated with the blue numbers to the right of the right axis; in units of $h^{-1}$Mpc) against $\ell$. On the top of each black solid curve the red number indicates the redshift of the corresponding $\chi$. The colour-coded lines along the vertical direction denote the values of $k$ that contributes at the given $\ell$ (as shown in the horizontal axis) and $\chi$ (as represented by the black solid curves); each of them corresponds to a fixed $k$ with the colour scheme as displayed in the colourbar on the top (running from black for $k\sim0.004~h$Mpc$^{-1}$ to red for $k=50~h$Mpc$^{-1}$). From Figure \ref{fig:decomp} we see the well-known result that most contributions at large (small) angular scales come from large (small) length scales; it also indicates that most contributions come from middle and low redshifts -- for example, $\sim90\%$ contributions in the case of $z_{\rm source}=1$ (0.5) come from the redshift range $z\lesssim0.7$ ($z\lesssim0.38$).

\begin{figure}
\includegraphics[width=\columnwidth]{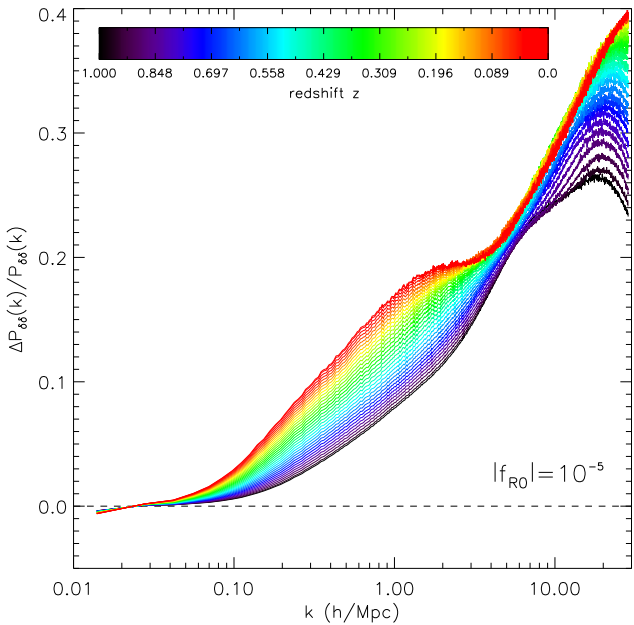}
\caption{(Colour Online) The relative enhancement of the nonlinear matter power spectrum $P_{\delta\delta}$ in F5 relative to GR, as a function of $k$ (the horizontal axis) and redshift $z(\chi)$ (equally spaced in $\chi$ with a $\Delta\chi=75h^{-1}$Mpc; colour-coded according to the colour-bar on the top).}
\label{fig:FR_PK}
\end{figure}

Fig.~\ref{fig:FR_PK} shows how the relative difference in the nonlinear $P_{\delta\delta}$ of F5 and GR evolves in time. We can see the known feature \citep[e.g.,][]{Li_PK_2013} that the F5 prediction agrees with GR on large scales due to the finite ranges of the modified gravitational force, and the enhancement of clustering on nonlinear scales ($k\geq0.1~h$Mpc$^{-1}$). We show 33 curves corresponding to the 33 snapshots between $z=1$ (black) and $z=0$ (red), which give us a comprehensive picture of not only the $k$- but also the time-dependences of the enhancement. For examples:

(i) At $k\sim0.1~h$~Mpc$^{-1}$, the relative enhancement of $P_{\delta\delta}$ for F5 is within $1-3$\% between $z=1$ and $z=0$. According to Fig.~\ref{fig:decomp} this $k$-mode is most relevant for $\ell\sim100$, which explains the $\sim2\%$ enhancement of $C_{\kappa\kappa}$ there (cf.~Fig.~\ref{fig:sim_halofit}, the $z_{\rm source}=1$ case -- the same below).

(ii) At $k\sim1~h$~Mpc$^{-1}$, $P_{\delta\delta}$ in F5 is enhanced by $10-18$\% between $z\sim0.7$ and $z=0$, leading to a $\sim15\%$ of $C_{\kappa\kappa}$ at $\ell\sim1000$.

(iii) Between $k=1$ and a few $h$Mpc$^{-1}$ is the transition regime between 1- and 2-halo terms in the halo model, where the enhancement in F5 $P_{\delta\delta}$ increases more slowly with $k$ than at lower $k$ values. This is reflected as a slight flattening of $\Delta C_{\kappa\kappa}/C_{\kappa\kappa}$ at $\ell$ just above $1000$ (Fig.~\ref{fig:sim_halofit}). 

(iv) At $k\sim5~h$~Mpc$^{-1}$, the relative enhancement of F5 $P_{\delta\delta}$ remains nearly a constant at $\sim20\%$ within $z=1$ and $z=0$, which translates into a $\sim20\%$ enhancement of F5 $C_{\kappa\kappa}$ at $\ell\sim4000-5000$ (Fig.~\ref{fig:sim_halofit}).

(v) At $k>5~h$~Mpc$^{-1}$, there is another steep increase of the enhancement in F5 $P_{\delta\delta}$ with $k$, indicating that dark matter haloes (at least the small ones which dominate in number) in F5 are more concentrated. This further increases $\Delta C_{\kappa\kappa}/C_{\kappa\kappa}$ above $\ell\sim4000-5000$.

(vi) At all $\ell$, $\Delta C_{\kappa\kappa}/C_{\kappa\kappa}$ is larger in the case of $z_{\rm source}=0.5$ than for $z_{\rm source}=1.0$, because in the former case the result is dominated by lower redshifts in the integration Eq.~(\ref{eq:pk2cl}), at which the enhancement of the nonlinear matter power spectrum in F5 relative to in GR is stronger.

The results here agree with those found in \citep{tessore15,shirasaki17}, and so serve as a sanity check of the simulations. Moreover, the decomposition of Figure~\ref{fig:decomp} provides a way to qualitatively understand the behaviour of the convergence power spectra based on knowledge about the scale- and time-dependences of the nonlinear matter power spectra. It can also prove useful when decomposing the degeneracies of modified gravity with other physical effects that can modify $P_{\delta\delta}$ on intermediate and small scales, such as Active Galactic Nuclei feedback and massive neutrinos \citep[see][for some general examples]{semboloni11,osato15,mummery17}; \citep[and see also e.g.,][for some examples in the framework of modified gravity]{arnold14,Harnois15}.

\subsection{Galaxy galaxy weak lensing}
\label{subsect:ggl}

Having looked at the behaviour of the lensing power spectra, now we move to the theoretical predictions of galaxy-galaxy weak lensing. 

\subsubsection{The galaxy matter cross correlation}

The correlation of shearing of source galaxies behind lens galaxies is given by the tangential shear profile around the lenses, which in turn is related to the excess surface density profile $\Delta\Sigma(r_p)$. This can be calculated by integrating $\xi_{gm}(r)$ along the l.o.s., and means that information about $\xi_{gm}$ can be obtained by studying galaxy galaxy lensing. For this reason in this subsubsection we shall compare the theoretical predictions of $\xi_{gm}$ by GR and F5.

We measure $\xi_{gm}$ directly by cross correlating the mock galaxy catalogues with the particle data from which they (and the corresponding halo catalogues) are constructed. This is done by a modified version of the the publicly available {\sc cute\_box} code \citep{cute}. We have not down-sampled the dark matter particles, and for a galaxy number of $N_g\sim3\times10^{5}$ and dark matter particle number $N_p=1024^3$ the code generally finishes for a snapshot in a few hours using 12 threads. For the auto-correlation function of matter, $\xi_{mm}$, the cost is prohibitive without down-sampling particles, and we instead calculate it by directly transforming the matter power spectra $P_{\delta\delta}$ (see below).

\begin{figure*}
\includegraphics[width=18.2cm]{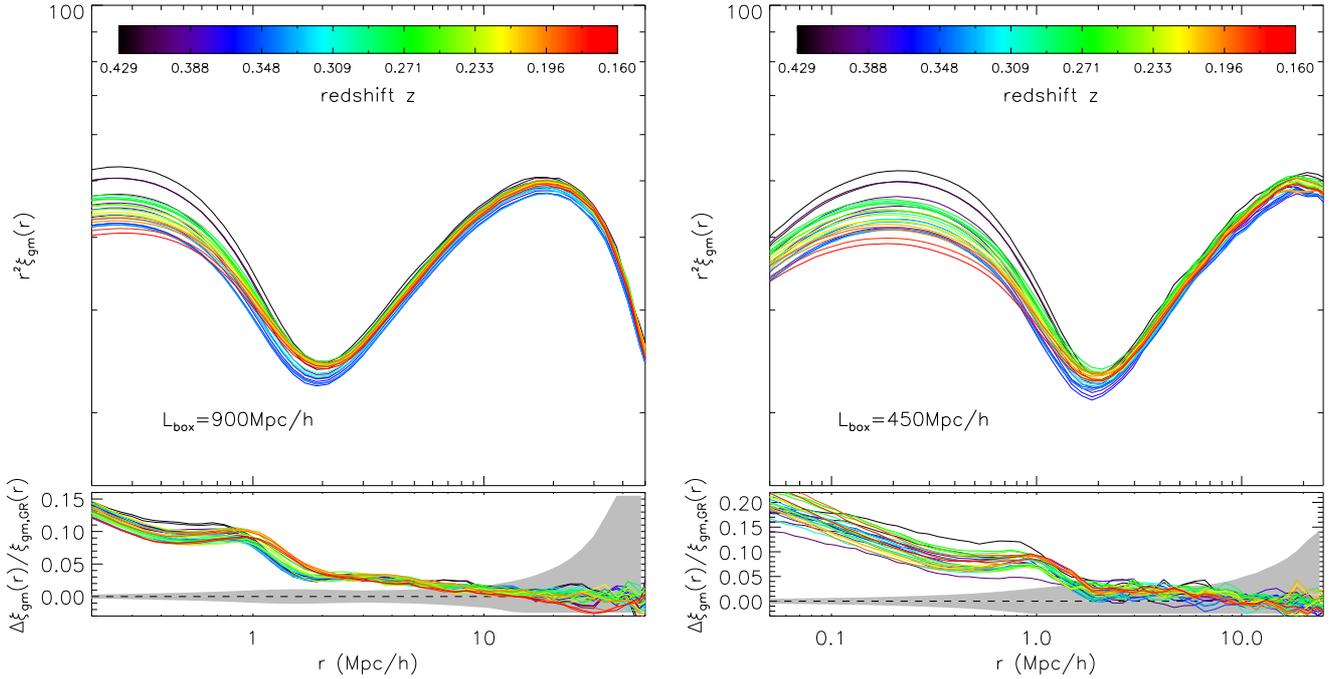}
\caption{(Colour Online) {\it Top left panel}: The galaxy-matter cross correlation function $\xi_{gm}(r)$, as a function of the separation $r$ of the galaxy-particle pair, for the 29 of HOD galaxy catalogues constructed using the GR $L_{\rm box}=900~h^{-1}$Mpc simulation. The colours of the curves indicate the redshift $z$ (see the color bar on the top). For clarify we have shown $r^2\xi_{gm}(r)$ and have not plotted the curves for F5. {\it Bottom left panel:} the relative differences between the F5 and GR predictions of $\xi_{rm}$ for the 29 HOD catalogues, using the same colour scheme. The grey shaded region marks the 1-$\sigma$ error from the 125 Jackknife resamples. {\it Right panels}: the same but for the simulation with $L_{\rm box}=450~h^{-1}$Mpc.}
\label{fig:xi_gm}
\end{figure*}

In order to estimate the uncertainties caused by the larger-scale environment, we use the internal jackknife resampling method with $N_{\rm JK}^3$ resamples by dividing the simulation box into $N_{\rm JK}^3$ subboxes of equal volume. For each resample, we discard the galaxies in one of the subboxes, but still include matter from that subbox when cross correlating with galaxies in other subboxes. The cross correlation function of the $i$-th jackknife resample, $\xi_{gm}^{(i)}$, is estimated as
\begin{equation}
1+{\xi}^{(i)}_{gm}(r,r+\Delta r) \equiv \frac{N^{(\bar{i})}_{\rm DD}(r,r+\Delta r)}{N_pN^{(\bar{i})}_g\Delta V/V} = \frac{N^{(\bar{i})}_{\rm DD}(r,r+\Delta r)}{n_pN^{(\bar{i})}_g\Delta V},
\end{equation}
where $V$ is the whole volume of the simulation box, $\Delta V$ is the volume of the radius bin $[r,r+\Delta r]$, $N_p$ the total number of dark matter particles in the simulation volume, $n_p=N_p/V$, and $N^{(\bar{i})}_g$ the number of galaxies and $N^{(\bar{i})}_{\rm DD}(r,r+\Delta r)$ the number of galaxy-particle pairs in radius bin $[r,r+\Delta r]$ when the $i$-th jackknife subbox is excluded. Note that we only discard the galaxies (and not the dark matter particles) in the $i$-th jackknife subbox when counting the pairs. We use $N_{\rm JK}=5$ in this work. Note that the jackknife error estimates do not include shape noises of source galaxies and the noises due to finite number of sources. Also, by directly integrating the galaxy-matter cross correlation function $\xi_{gm}$, we essentially perform the integration along all possible lines of sight and so the statistical error is smaller than that obtained using a sample of the same volume and number density of lens objects in real observations, where one has only a single line of sight for each object. For a more detailed analysis of errors using mock lensing data, see \citet{shirasaki16}.

Figure \ref{fig:xi_gm} presents the galaxy matter cross correlation functions measured from the $L_{\rm box}=900$ (left) and $450~h^{-1}$Mpc (right) simulations, and for each case we show, in the top panel, the GR results for all 29 snapshots of HOD catalogues (the coloured solid curves, with the colours denoting the redshifts according to the colour bar). In the lower panels we show the relative differences between F5 and GR using the same redshift-colour scheme, and the gray shaded region marks the 1-$\sigma$ Jackknife error. Notice that to improve visibility we have plotted $r^2\xi_{gm}(r)$ instead of $\xi_{gm}(r)$ itself in the top panels. The results from the two boxes agree with each other well.

The GR curves in Fig.~\ref{fig:xi_gm} show a scatter of up to $\sim10\%$, which is because the different HOD catalogues do not correspond to the same set of tracers of the density field. To see this more explicitly, let us remark that the galaxy number density in our HOD catalogues peaks at $z\sim0.16$, and as we go to higher redshifts it first decreases until $z\sim0.27$, when it begins to increase, and then from $z\sim0.33$ it starts to decrease again. The same trend can be seen in the $\xi_{gm}$ curves of Figure~\ref{fig:xi_gm}, namely the curves become higher from $z\sim0.16$ (red) to $0.27$ (green), then lower between $z\sim0.27$ and $z\sim0.33$ (blue), and then higher again until $z\sim0.43$ (black). The trend is the same from both boxes. Physically, according to our HOD model, a lower galaxy number density indicates that only the more massive haloes are populated with galaxies, and these haloes have stronger correlations with matter around and inside. The behaviour is hence not surprising: when we consider the cross correlation between the matter density field and its tracers, the choice of the latter is critical (we will revisit this point later).

What is more interesting is the relative difference between F5 and GR, which is shown in the lower panels of Figure \ref{fig:xi_gm}. There we see that the two models agree with each other (within the Jackknife error) at $r>10~h^{-1}$Mpc, that F5 $\xi_{gm}$ starts to get enhanced -- with a modest enhancement factor of $3$-$5$\% -- relative to the GR result at $r<10~h^{-1}$Mpc, and finally near the transition scale between 1- and 2-halo terms ($r\sim1$-$2~h^{-1}$Mpc) the enhancement factor increases to $\sim10\%$. The stronger enhancement in the regime where the 1-halo term dominates is related to the more concentrated matter distribution in F5 compared to GR. Note that the model difference, unlike the cross correlation function $\xi_{gm}$ itself, is less dependent on redshift and therefore less sensitive to the choice of tracers. This is also as expected, because the selection of tracers in our HOD modelling affects both models in a similar way.

\subsubsection{The excess surface density profiles}

\begin{figure*}
\includegraphics[width=18.2cm]{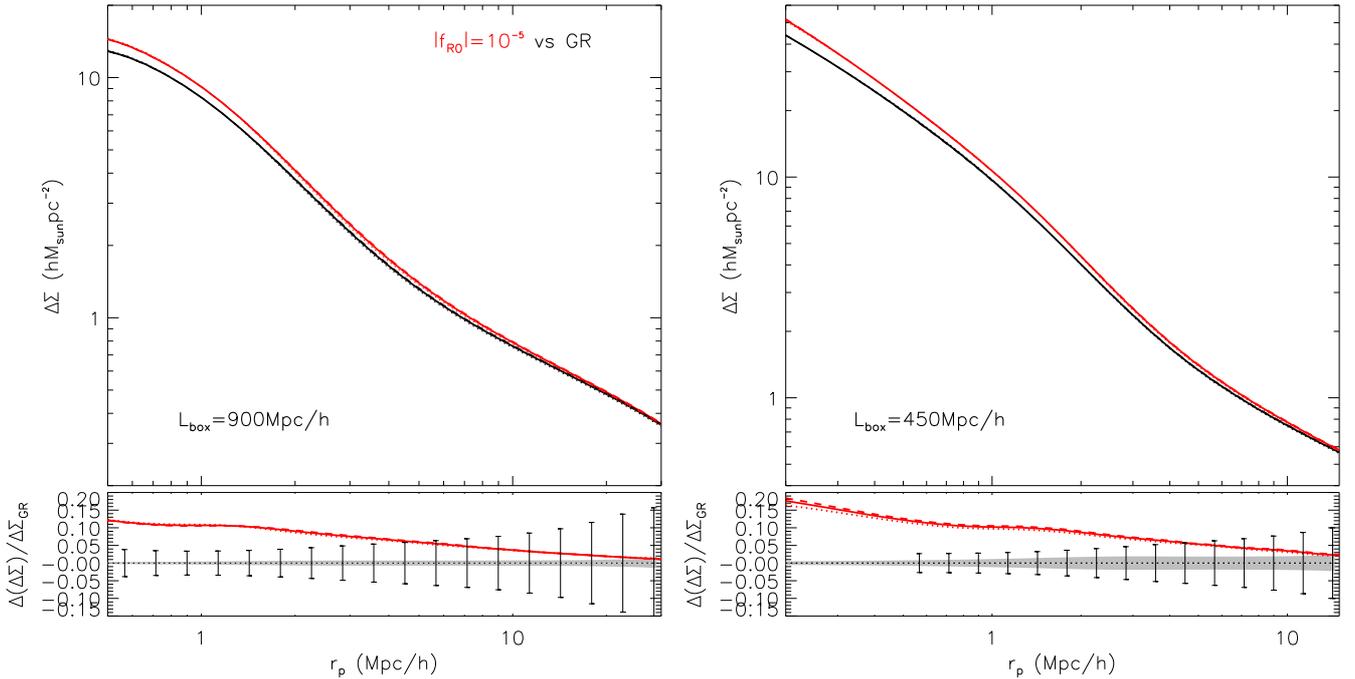}
\caption{(Colour Online) {\it Top left panel}: The excess surface density profiles as functions of the projected separation $r_p$, measure from the $L_{\it box}=900h^{-1}$Mpc GR (black solid line) and F5 (red solid line) simulations, assuming a stack around all galaxies in the 29 snapshots of HOD galaxies. The dotted and dashed lines -- with the same colour -- are the results of using the 14 high-redshift ($0.30\lesssim z\lesssim0.43$) and 15 low-redshift ($0.16\lesssim z\lesssim0.29$) HOD catalogues, and they are barely visible in the plot. {\it Bottom left panel}: The relative difference between F5 and GR, using the same line styles and colours (now the difference between the high-$z$ and low-$z$ galaxy samples is clearer but still very small). The gray shaded region shows the square root of the diagonal elements of the Jackknife covariance matrix, and the error bars are the square root of the diagonal elements of the analytical covariance matrix assuming the {\sc hsc} survey. {\it Right panels}: the same but for the simulation with $L_{\rm box}=450~h^{-1}$Mpc.}
\label{fig:Delta_Sigma}
\end{figure*}

The averaged excess surface density profile, $\Delta\Sigma(r_p)$, is calculated by directly integrating the galaxy-mass cross correlation functions, following the prescription of Eq.~(\ref{eq:excess}). 

To calculate this, we first use cubic spline to interpolate the $\xi_{gm}\left(r=\sqrt{r^2_p+z^2}\right)$ measured from the simulations onto a grid in $\left({\rm log}(r_p), z\right)$, where $r_p$ is the projected separation and $z$ is the l.o.s. coordinate (with $z=0$ at the position of the galaxy)\footnote{This is a slight abuse of notation, but it should be clear, given the context, where $z$ means the l.o.s.~coordinate or redshift.}, and then perform the integrations using direction summation. We have chosen a fine interpolation grid, as well as a large enough $z_{\rm max}$ for the los~integration, so that the percentage error for the numerical integrations is within $\sim0.1\%$. For the $L_{\rm box}=900~h^{-1}$Mpc ($450~h^{-1}$Mpc) simulations, this allows us to have $\Delta\Sigma(r_p)$ up to $r_p=30~h^{-1}$Mpc ($15~h^{-1}$Mpc), with $z_{\rm max}=90~h^{-1}$Mpc ($\sim65~h^{-1}$Mpc).

The left (right) panel of Figure \ref{fig:Delta_Sigma} shows the results of $\Delta\Sigma(r_p)$ from the $L_{\rm box}=900~(450)~h^{-1}$Mpc simulation. A comparison of the upper panels indicates that the $900~h^{-1}$Mpc box starts to lose resolution at $r_p\sim1~h^{-1}$Mpc, but the relative difference between F5 and GR agrees down to $r_p\sim0.5~h^{-1}$Mpc. In both cases, we have split the lens galaxies into two separate redshift ranges (a high-$z$ bin with $0.30\lesssim z\lesssim0.43$ and a low-$z$ bin with $0.16\lesssim z\lesssim0.29$), but the results -- dotted and dashed lines respectively -- display barely any difference from using the whole lens galaxy sample, suggesting that the conclusion does not depend sensitively on redshift or the source number density, at least in the redshift range covered in this study.

The stacked excess surface density profiles in F5 and GR differ most significantly at $r_p\lesssim2~h^{-1}$Mpc, which reflects the bigger difference in the 1-halo term of the matter-galaxy correlation function as shown in Fig.~\ref{fig:xi_gm}. The model difference between F5 and GR has a distinct dependence on $r_p$ from that of $\xi_{gm}(r)$, which is $\sim5$-$10\%$ at $1\lesssim r_p\lesssim10~h^{-1}$Mpc, because $\Delta\Sigma(r_p)$ depends on the average $\xi_{gm}$ from $R=0$ to $R=r_p$, where $R$ is the projected distance from the lens. Although this means that one can see a strong signal up to larger values of $r_p$, it also implies that uncertainties that affect the prediction of the 1-halo term in $\xi_{gm}$, such as baryonic physics, can have an impact on $\Delta\Sigma$ to larger $r_p$, something which we should bear in mind when considering observational constraints.

\begin{figure*}
\includegraphics[width=18.2cm]{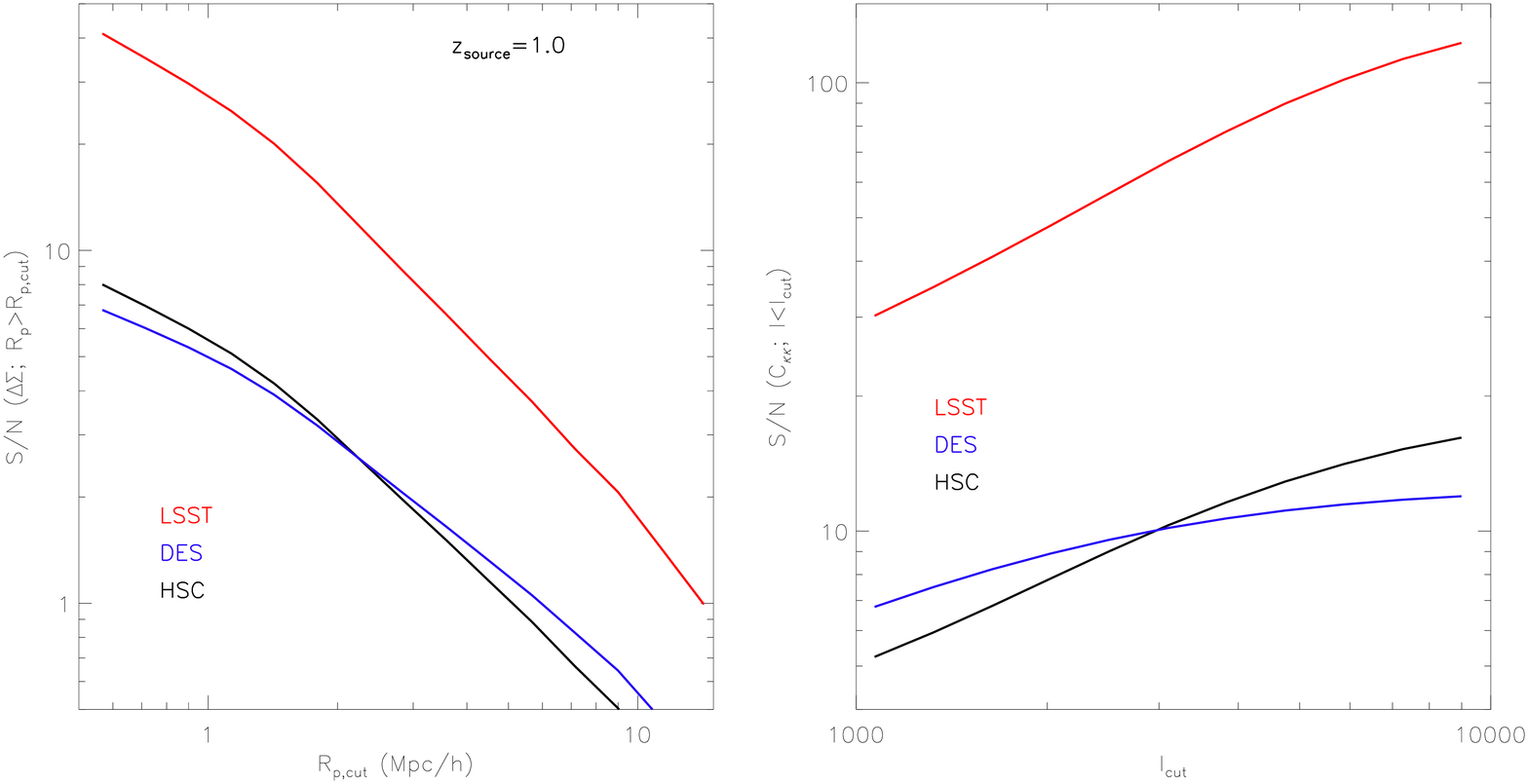}
\caption{(Colour Online) {\it Left panel}: the S/N, defined in Eq.~(\ref{eq:SN}), from galaxy-galaxy lensing as a function of $R_{p,{\rm cut}}$, which is the minimum projected radius $R_p$ used in the calculation. $R_{p,{\rm max}}=25~h^{-1}$Mpc here and the model differences are calculated using the results from the $L_{\rm box}=450~h^{-1}$Mpc simulations. The black, blue and red curves are the results by using the covariance matrices for the assumed {\sc hsc}, {\sc des} and {\sc lsst} surveys respectively. {\it Right panel}: the S/N from the lensing convergence power spectrum as a function of the maximum $\ell$ ($\ell_{\rm cut}$) used in the analysis, with $\ell_{\rm min}=200$. The colours of the curves are the same for the left panel.}
\label{fig:SN}
\end{figure*}

\begin{table}
\caption{The three hypothetical lensing surveys used to estimate the statistic error for galaxy-galaxy weak lensing and the convergence power spectrum. $n_{\rm gal}$ is the source galaxy number density (the sources are assumed to be at $z_{\rm source}=1.0$ in all cases), and $\sigma_{\rm int}$ the intrinsic ellipticity of sources. For galaxy-galaxy lensing we have assumed the lens galaxies are at $z_{\rm lens}=0.30$.}
\begin{tabular}{@{}llll}
\hline\hline
Hypothetic Survey & $n_{\rm gal}$ (${\rm arcmin}^{-2}$) & $\sigma_{\rm int}$ & Survey Area (${\rm deg}^2$) \\
\hline
{\sc des}  & $3$ & $0.35$ & $5000$\\
{\sc hsc}  & $30$ & $0.35$ & $1400$\\
{\sc lsst}  & $50$ & $0.35$ & $20000$\\
\hline
\hline
\end{tabular}
\label{table:survey}
\end{table}

The $10$-$15$\% relative difference in $\Delta\Sigma$ at $r_p\lesssim1h^{-1}$Mpc between the two models is roughly the same as the difference in $C_{\kappa\kappa}$ at $\ell\lesssim2000$, but smaller than the difference in $C_{\kappa\kappa}$ at larger $\ell$ (see Fig.~\ref{fig:sim_halofit}), and therefore we are interested in whether there is any difference in the signa-to-noise (S/N) of these two probes. The S/N quantifies the distinguishability of the two models and is defined as 
\begin{equation}\label{eq:SN}
\left({\rm S/N}\right)^2 \equiv \left[d_{\rm F5}(x_i)-d_{\rm GR}(x_i)\right]^{\rm T}{\bf C}^{-1}(x_i,x_j)\left[d_{\rm F5}(x_j)-d_{\rm GR}(x_j)\right],
\end{equation}
in which $x_i$ is the data in the $i$th bin ($x=R_p$ for GGL and $x=\ell$ for lensing convergence), and ${\bf C}$ is the covariance matrix.

To estimate the S/N values, we consider three fiducial surveys with roughly the {\sc des}, {\sc hsc} and {\sc lsst} specifications, as summarised in Table \ref{table:survey}. We expect that Euclid will have a similar performance as {\sc lsst} \citep[see, e.g.,][for some relevant S/N analysis but for void lensing using Euclid and {\sc lsst}]{cautun17}. For simplicity, let us assume that there are overlapping spectroscopic surveys which can provide galaxy catalogues with a number density at least as high as the ones used in our lensing galaxies, between $0.16\lesssim z\lesssim0.43$. As shown above, the model difference in $\Delta\Sigma(r_p)$ is insensitive the lens redshift, so we assume a single lens redshift $z_{\rm lens}=0.3$ and a single source redshift $z_{\rm source}=1$ in the simplified study. The covariance matrix ${\bf C}$ is calculated based on a halo model prescription following \citet{jeong09}, which accounts for contributions from cosmic variance, the Poisson noise of lensing galaxies and the shape noise of source galaxies, and it adopts single source and lens redshifts as was mentioned above. The GR HOD and cosmological parameters for the snapshot at $z\sim0.3$ are used to generate ${\bf C}$. 
We also adopt the halo-model approach developed in \citet{Sato09} to estimate the covariance matrix for $C_{\kappa\kappa}$. In this matrix, we properly include the non-Gaussian term induced by convergence tri-spectrum and halo sample variance.

Figure \ref{fig:SN} shows the S/N based on the our analysis which takes into account statistical uncertainties only, for the GGL (left panel) and $C_{\kappa\kappa}$ (right panel). The black, blue and red curve are respectively for a {\sc hsc}, {\sc des} and {\sc lsst}-like survey. In the case of GGL we show S/N as a function of the minimum $r_p$ used in the calculation, while for the case of $C_{\kappa\kappa}$ the S/N is displayed as a function of the maximum $\ell$ included in the analysis. These are because using a smaller (larger) cutoff in $r_p$ ($\ell$) means more data are included in the model test, which can increase the S/N; but on the other hand, by doing these we are moving to smaller scales, where we would need to worry about other theoretical uncertainties such as the impact of baryons. Eventually, a comprise between these two considerations will have to be made; but for this study we are mainly interested in how the S/N values vary with the $r_p$ and $\ell$ cuts.

From Fig.~\ref{fig:SN} we can see that {\sc lsst}, being a stage-4 experiment, has significantly more power to distinguish the two models, thanks to its larger sky coverage and higher source galaxy number density; this is true for both GGL and $C_{\kappa\kappa}$. {\sc hsc} and {\sc des}, both of which are stage-3 experiments, show similar performances. When comparing the two probes, we find that $C_{\kappa\kappa}$ gives larger S/N values for all 3 surveys -- this is not surprising as both can be symbolically written as $\langle AB\rangle$ where $A, B$ are two statistics of the underlying density field and $\langle\cdot\rangle$ denotes ensemble average. For $C_{\kappa\kappa}$, $A=B=\kappa$ while for GGL we have $A=\kappa$ and $B=\delta n_g$, where $\delta n_g$ is the galaxy number density fluctuation, the model difference in which has been greatly reduced by the tuning of HOD parameters. Therefore, we expect the model difference in GGL to be smaller than that in $C_{\kappa\kappa}$.

\subsubsection{Effect of tracers}

\begin{figure}
\includegraphics[width=\columnwidth]{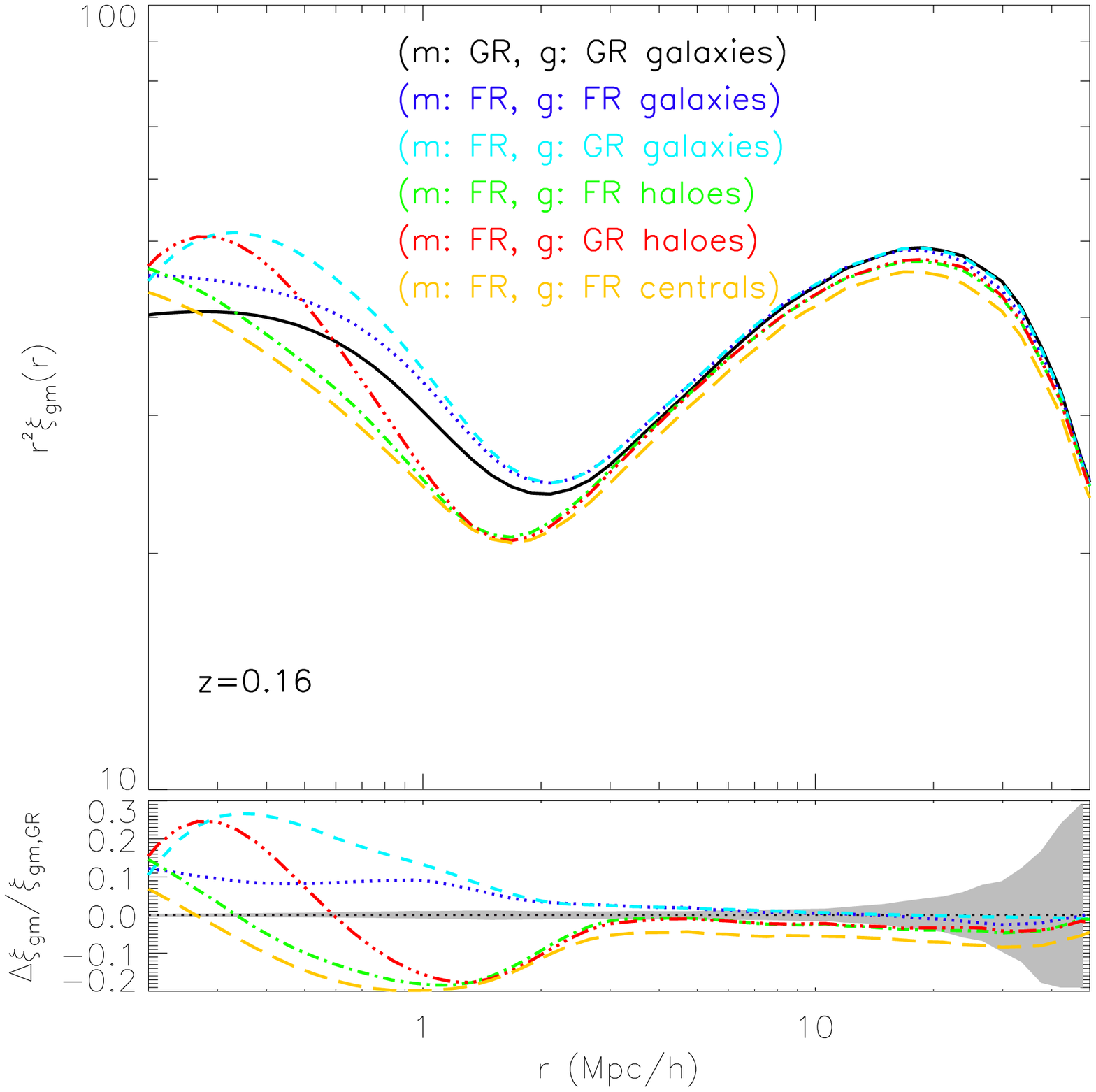}
\caption{(Colour Online) {\it Top Panel}: $\xi_{gm}$ for different choices of tracers to cross correlate with matter. The black solid curve is the cross correlation between HOD galaxies and dark matter both from the GR $L_{\rm box}=900h^{-1}$Mpc simulation, while the other curves are cross correlations of dark matter from the F5 simulation with a variety of choices of tracers -- HOD galaxies form the F5 simulation (blue dotted), HOD galaxies for GR (cyan dashed), dark matter haloes for F5 (green dot-dashed), haloes for GR (red dot-dot-dot-dashed) and central HOD galaxies for F5 (orange long dashed). In the cases of haloes being the tracers, the haloes are ranked from high to low mass and the first $\bar{n}_{g,{\rm GR}}L^3_{\rm box}$, where $\bar{n}_{g,{\rm GR}}$ is the averaged HOD galaxy number density from the GR simulation at $z=0.16$, are used. {\it Bottom Panel}: The relative differences between the other cases and the $\xi_{gm}$ for GR galaxies and dark matter, with the same line styles and colours as in the top panel. The yellow shaded region shows the 68\% scatter of the 125 Jackknife resamples around the mean, for the cross correlation between galaxies and dark matter in GR.}
\label{fig:xi_cross_checks}
\end{figure}

Because the galaxy-matter cross correlation quantifies the distribution of matter around galaxies, the choice of galaxies is important, which is why we made effort to tune the HOD parameters in the F5 case to match the 3D galaxy two point correlation function $\xi_{gg}$ (and the galaxy number density) in the two models. To gain a feeling of the effect of choosing different tracers of the matter field, we have, for the snapshot at $z=0.16$\footnote{We only use one snapshot because, as we have seen above, the model difference between F5 and GR is shows little redshift evolution in the redshift range considered in this paper.} of the $L_{\rm box}=900h^{-1}$Mpc simulation, tested how $\xi_{gm}$ changes by using other tracers. The result is shown in Figure~\ref{fig:xi_cross_checks} (see the figure caption for a more detailed description of the different curves). There are several interesting features in this figure.

First of all, we see that at small separations $\xi_{gm}$ is indeed very sensitive to the choice of tracers, as expected. The difference caused by this is substantially larger than that between F5 and GR as we have show above. Evidently, for the one-halo term of $\xi_{gm}$, changing the tracer is equivalent to placing the tracer to a different location in its host halo, or even outside the halo, which can impact the matter distribution around. This sensitivity is reduced at large separations, at which the effect of slightly relocating the tracer is smaller. 

Second, for the two cases of cross correlating dark matter distribution from the F5 simulation with haloes (green dot-dashed and red dot-dot-dot-dashed), the lowest points in the curve shift toward smaller separations, and $\xi_{gm}$ is overall lower than the cross correlations by using HOD galaxies. Using haloes as tracers is similar to using only the central galaxies, and misses the contribution from satellite-matter cross correlation, which may have caused this feature. To check this, we show the result of cross correlating the dark matter field with only the central HOD galaxies, in the case of F5, as the orange long dashed curve in Figure \ref{fig:xi_cross_checks}. The behaviour is very similar to the cross correlation with haloes, and the slight difference is because the two tracer classes do not correspond to exactly the same halo population since central galaxies are populated in haloes in a random way.

Third, for the two cases in which we cross correlate the dark matter field from the F5 simulation with tracers from the GR simulation (cyan dashed and red dot-dot-dot-dashed), $\xi_{gm}$ decreases toward very small separations ($r\lesssim0.3h^{-1}$Mpc), and this is because GR tracers usually do not coincide with the highest density peaks in the dark matter field of the F5 simulation so that the highest values of $\xi_{gm}$ appear to be away from the GR tracers themselves.

All in all, the test highlights the importance of using carefully constructed mock galaxy catalogues when we compare the predictions of galaxy-galaxy lensing in different models. As an example, the 3D real-space two point correlation functions and the projected 2D correlation functions in our GR and F5 HOD catalogues agree with each other within $2$-$3\%$ in the range of separations considered here, but such agreements are in a statistical, rather than an object-by-object, sense. When cross correlated with the same (F5) matter field, the $\xi_{gm}$ results by using F5 and GR galaxies are indeed different (cf.~blue dotted and cyan dashed curves in Fig.~\ref{fig:xi_cross_checks}), in particular at small separations. This indicates that high-resolution simulations enable to accurately resolve the internal structures of haloes and to allow realistic galaxy mocks to be constructed, are necessary in order to make galaxy-galaxy lensing more useful for distinguishing $f(R)$ gravity and GR.


\subsection{Galaxy bias}
\label{subsect:bias}

Finally, we compare the galaxy biases in GR and $f(R)$ gravity. We are interested in this comparison this for two reasons. 

Firstly, in standard $\Lambda$CDM, the linear bias factor, defined as
\begin{equation}
b_{\rm lin}(r) = \frac{\xi_{gm}(r)}{\xi_{mm}(r)} = \frac{\xi_{gg}(r)}{\xi_{gm}(r)} = \sqrt{\frac{\xi_{gg}(r)}{\xi_{mm}(r)}},
\end{equation}
is scale-independent on large scales. On the other hand, it is known that $f(R)$ gravity, and in general chameleon models, predict a scale-dependent linear growth rate of matter density perturbations, which may result in scale-dependent $b_{\rm lin}$, because we have tuned the HOD parameters to match $\xi_{gg}(r)$ in the two models, while their $\xi_{mm}(r)$ can have a scale-dependent difference. We want to check whether, at least for F5, such a scale dependence of $b_{\rm lin}$ is significant enough to make it of interest in observations.

Secondly, assuming that the linear bias factor $b_{\rm lin}$ gives a correct quantification of galaxy bias at some given scale $r$, the galaxy-matter cross correlation coefficient $R_{gm}$, defined as
\begin{equation}
R_{gm}(r) \equiv \xi_{gm}(r)/\sqrt{\xi_{gg}(r)\xi_{mm}(r)}, 
\end{equation}
is equal to 1. This offers an opportunity to derive the matter correlation function $\xi_{mm}$ from observational determinations of $\xi_{gg}$ and $\xi_{gm}$. We want to see whether the minimum length scale $R_{\rm min}$ above which this can be performed without worrying about the nonlinear effects in galaxy bias is significantly different in GR and in F5. 

\begin{figure*}
\includegraphics[width=18.2cm]{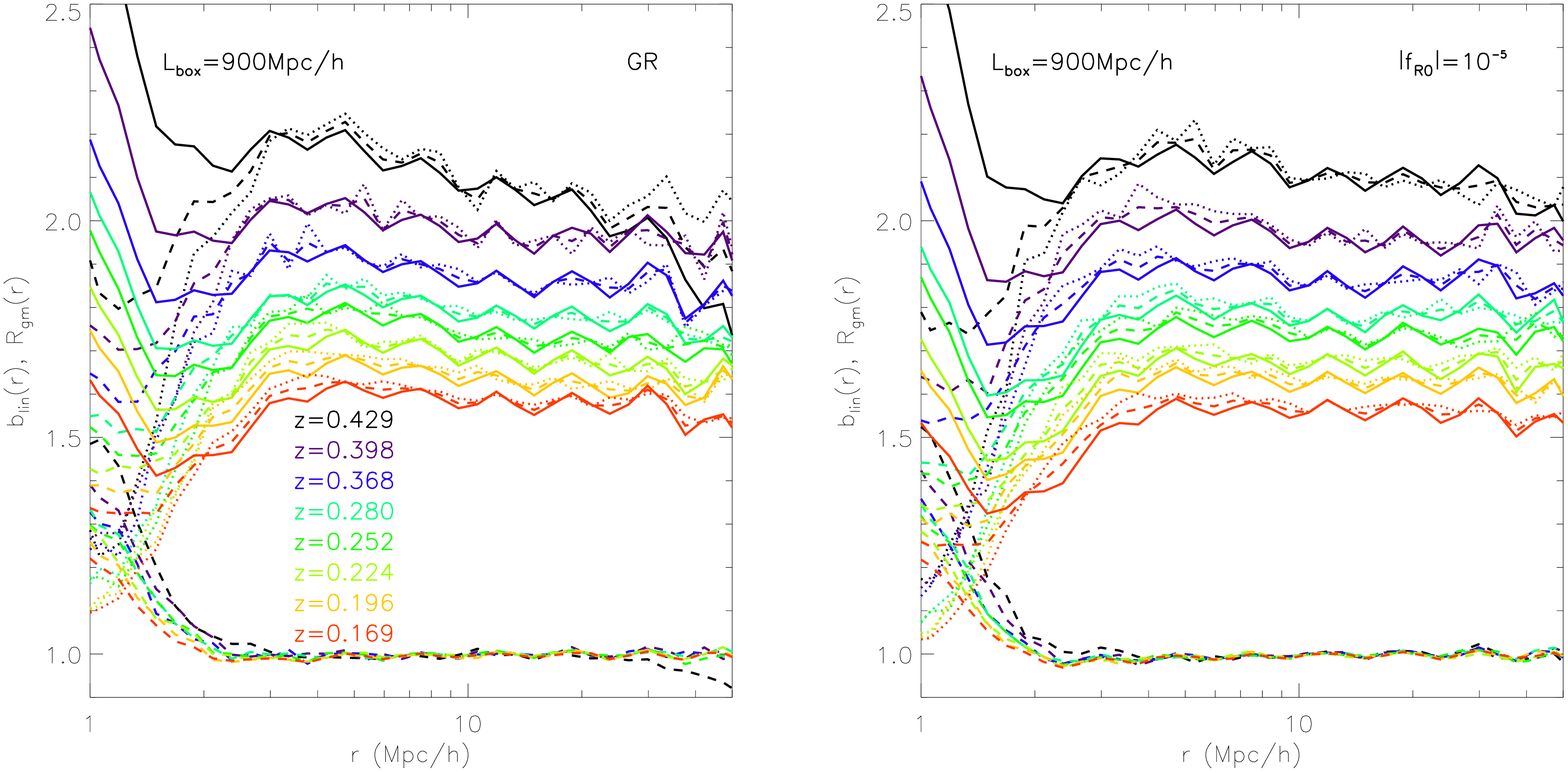}
\caption{(Colour Online) {\it Left panel}: The bunch of curves on the top of the panel show the linear galaxy bias defined in three ways -- $b_{\rm lin}\equiv\xi_{gm}/\xi_{mm}$ (solid), $b_{\rm lin}\equiv\sqrt{\xi_{gg}/\xi_{mm}}$ (dashed) and $b_{\rm lin}\equiv\xi_{gg}/\xi_{gm}$ (dotted) -- from the GR $L_{\rm box}=900~h^{-1}$Mpc simulation. The redshifts of the curves are given in the legend, using the same colour scheme and order (from top to bottom) as the curves themselves. The dashed lines at around $1.0$ are the galaxy matter correlation coefficient defined as $R_{gm}\equiv\xi_{gm}/\sqrt{\xi_{gg}\xi_{mm}}$ for the same redshifts. {\it Right panel}: The same as the left panel, but for F5.}
\label{fig:bias}
\end{figure*}

Figure \ref{fig:bias} shows the results of our tests to answer these questions. The top bunch of curves in each panel (left for GR and right for F5) show the linear bias factors calculated in different ways (the different line styles) at a selection of redshifts (shown by different colours; see the legend for more details). We see that the different ways of calculating $b_{\rm lin}$ agree well with each other, and that $b_{\rm lin}$ remains roughly constant down to $r\sim4~h^{-1}$Mpc, in agreement with previous results \citep[e.g.,][]{yoo2006}. This holds true for all redshifts and both models -- the latter is perhaps not surprising given that $\xi_{gg}$ has been tuned to match in GR and F5, while $\xi_{gm}$ in these models differ by at most a few percent above $r\sim4~h^{-1}$Mpc, so that any scale dependence of $b_{\rm lin}$ should be very weak. Note that $b_{\rm lin}$ is larger at higher redshifts since the number densities of galaxies are lower for those redshifts and so we are looking at more biased tracers of the underlying matter field.

The results of $R_{gm}$ are shown as dashed lines near the bottom of the left (right) panel for GR (F5), for the same redshifts and using the same colour scheme. We can see that for $r$ down to $1$-$2~h^{-1}$Mpc $R_{gm}$ is equal to unity to a good accuracy, and there is no significant difference between the different redshifts or between GR and F5. This suggests that the reconstruction of $\xi_{mm}$ from $\xi_{gg}$ and $\xi_{gm}$ can be done with the same reliability whether the underlying model is GR or F5 -- this is even with the consideration that $b_{\rm lin, F5}$ may vary by a few percent (in theory) from vary large scales to $r\sim4~h^{-1}$Mpc. 

\section{Summary and discussion}
\label{sect:con}

We have studied the possibility of using galaxy-galaxy weak lensing to distinguish between the standard $\Lambda$CDM paradigm and a lead alternative, chameleon $f(R)$ gravity. The parameter for the latter is chosen so that the model is not yet firmly ruled out by cosmological data. The study can also lead to a more quantitative assessment of to what accuracy deviations -- in the way as prescribed by chameleon type theories -- from GR on cosmological scales can be constrained by future GGL observations. To this end, we have decided to make accurate theoretical predictions for a representative MG model, by the means of fully nonlinear N-body simulations.

We take a step further by carefully making appropriate galaxy catalogues in order to predict the GGL signals for both $\Lambda$CDM and $f(R)$ gravity. The reason is this: while the $f(R)$ model studied here has a stronger gravitational interaction than in GR, and thus would predict more dark matter haloes and an enhanced large-scale clustering,  the lack of a reliable predictive model for galaxy-halo connection means that we can not rule out it on the basis of its predicted galaxy clustering using some approximate method. Hence, we follow a more pragmatic approach, by assuming that both models produce an acceptable galaxy clustering (as required by observations), and focusing only on the resulting lensing effect. In practice, this is achieved by adopting the HOD prescription of galaxy-halo connection, and tuning the HOD parameters in the $f(R)$ model to match the galaxy clustering in its $\Lambda$CDM counterpart. We have checked explicitly, e.g., in Fig.~\ref{fig:xi_cross_checks}, that the choice of tracers (the galaxy catalogues) has a non-negligible effect on the resulting GGL signal.

Before looking at GGL, we have first inspected the weak lensing convergence power spectrum in the selected $f(R)$ model. This has been studied before, but here we focus on the connection to the matter power spectrum and its time evolution. Figure \ref{fig:decomp} is a useful plot to understand how much contribution the convergence power spectrum has received from the matter clustering at different scales and epochs. We have checked that the behaviour of the former can be qualitatively explained using the latter.

Compared to the convergence and matter power spectra, which quantify the auto-correlation of the matter density field, we find that the cross correlation between galaxies and matter, $\xi_{gm}(r)$, shows a $\sim50\%$ smaller difference between the two model studied, because the galaxy clustering in these models has been tuned to agree well with each other. This is further verified by an analysis of the signal-to-noise (S/N), which quantifies the distinguishability of the models when statistical-only errors are used, for three imaging surveys similar respectively to the {\sc des}, {\sc hsc} and {\sc lsst} specifications, with synergy data from a spectroscopic survey that can at least match the BOSS LOWZ galaxy number density at $0.16\lesssim z\lesssim0.43$. We find that the S/N is smaller than $2\sim3$ for {\sc des} and {\sc hsc}, if we exclude GGL data within $2\sim3h^{-1}$Mpc, while for the {\sc lsst}-like survey we have ${\rm S/N}\sim10$. The S/N is $2\sim4$ times higher if the cosmic shear power spectrum, with a conservative cut at $\ell_{\rm min}=1000\sim2000$, is used for constraints. In the redshift range covered in this study, we do not find any strong dependence of the model difference in GGL on the lens redshift -- indeed, Fig.~\ref{fig:Delta_Sigma} shows that the low-$z$ and high-$z$ samples give almost identical results.

Note that in the above forecast we have included only statistical and not systematic uncertainties. {Although the cosmic shear power spectrum can have a higher S/N than GGL for distinguishing $f(R)$ gravity, it still remains useful and complementary to consider GGL. Current two-point correlation analysis of cosmic shear is subject to observational systematics induced by imperfect measurement of distant galaxies. The multiplicative bias is one of the most important systematics in cosmic shear analysis and can make the amplitude of lensing power spectrum uncertain \citep[e.g.][]{Huterer06}. In recent cosmological analyses, the impact has been controlled  by image simulations \citep[e.g.,][]{kids450} or appropriate parametralisation \citep[e.g.][]{DESY1}. On the other hand, GGL analysis can separate the multiplicative bias and lensing effects from observed signal if one can use single source population and multiple foreground objects at different redshifts \citep{oguri11}. We also note that the intrinsic alignments of galaxy ellipticities, which are a significant source of systematic errors for cosmic shear measurements \citep[e.g.][]{hirata04}, are not important for GGL with single source population. This is because the stacked lensing is linear in shear, and does not include any correlation between the shapes of different galaxies.}

We have investigated galaxy bias in the $\Lambda$CDM and $f(R)$ models. It is well known that $f(R)$ gravity models have scale-dependent linear growth rate, that goes back to the $\Lambda$CDM prediction on large scales. Figure~\ref{fig:FR_PK} is an example of this scale dependence, although it shows the {\it non}linear growth of matter density perturbations. Hence, we expect the linear galaxy bias to be different in these two models. However, from Fig.~\ref{fig:bias} we do not find a clear difference: the different values at the different redshifts are more likely due to the different corresponding galaxy number densities, as we know that the more luminous galaxies (for which $n_g$ is smaller) are more strongly biased tracers of the dark matter field. The F5 galaxy bias, at a given redshift, only shows a rather weak scale dependence and is slightly smaller than the GR value, the latter being because $\xi_{gg}$ is tuned to agree in the two models while $\xi_{mm}$ is larger in $f(R)$ gravity. On the other hand, we find that, as in GR, the galaxy-mass cross correlation coefficient $R_{gm}$ remains at $1.0$ down to $\sim3h^{-1}{\rm Mpc}$, suggesting that one can infer $\xi_{mm}$ from measurements of galaxy auto correlation and GGL at scales larger than $\sim3h^{-1}{\rm Mpc}$.

As a first study of GGL in $f(R)$ gravity, this work can be further improved in several ways, all of which require more (and more advanced) simulations than used here. 

First, as shown in Fig.~\ref{fig:xi_gm}, the galaxy-mass cross correlation $\xi_{gm}$ (and therefore GGL) depends sensitively on  the number density of lenses. Our galaxy catalogues are made to mimic the BOSS LOWZ sample, with a number density of order $10^{-4}\left(h^{-1}{\rm Mpc}\right)^{-3}$, and this number will be further increased with ongoing and future galaxy surveys. It remains interesting to see how effectively the increased lens number density can help improve the S/N, given that many of the newly added galaxies will be fainter. In order to check this, we will need simulations with higher resolutions to populate fainter galaxies, and such simulations may also enable mock galaxy catalogues constructed using other techniques such as SHAM and SAM, to understand the impact of different methods of galaxy-halo connection.

Second, the simulations used here are for a choice of particular cosmological parameters (WMAP9). It is important to understand whether other choice of cosmological parameters can have a significant impact on the model difference in GGL. It is also important to check if there is a degeneracy between modified gravity and cosmology, e.g., whether the effect of $f(R)$ gravity can be mimicked by a different value of $\sigma_8$ or $\Omega_m$, as such a degeneracy can downgrade the potential of using GGL to constraint chameleon $f(R)$ models. For example, it was shown \citep{yoo2006} that the GGL signal $\Delta\Sigma(r_p)\propto\Omega_m^\alpha\sigma_8^\beta$, with $\alpha, \beta$ depending weakly on $r_p$, and therefore it is possible that the $r_p$-dependence of $\Delta(\Delta\Sigma)/\Delta\Sigma_{\rm GR}$ is not completely degenerate with the effects of varying $\Omega_m$ and $\sigma_8$, allowing these different parameters to be constrained simultaneously using GGL alone, or possible synergies of GGL with other probes. This will require simulations with different cosmologies. 

Third, in the forecast above we have deliberately not used signals on smaller scales ($R_r\lesssim2\sim3h^{-1}{\rm Mpc}$ and $\ell>1000\sim2000$), not just for resolution consideration, but also to be conservative as on those scales the theoretical predictions might be significantly affected by poorly-understood baryonic physics (which is not included in the simulations used in this work). The modified gravitational force can change the halo density profiles, which is reflected in the GGL signals. However, to reliably include this in the model test, we have to fully understand how baryonic processes affect the (re)distribution of matter inside dark matter haloes for $f(R)$ gravity, and for this hydrodynamical simulations with suitably adjusted subgrid physics are essential. This is an almost entirely unexplored regime to date.

\section*{Acknowledgements}

We would like to thank Christian Arnold, Alex Barreira, Carlton Baugh, Yan-Chuan Cai, Marius Cautun, Jianhua He, Peder Norberg for useful discussions. The authors thank the kind host by Naoki Yoshida at IPMU, the University of Tokyo, when part of this work was carried out. BL is supported by an European Research Council Starting Grant (ERC-StG-716532-PUNCA) and UK Science and Technology Facilities Council (STFC) Consolidated Grants (ST/P000541/1 and ST/L00075X/1). This work used the DiRAC Data Centric system at Durham University, operated by the Institute for Computational Cosmology on behalf of the STFC DiRAC HPC Facility (www.dirac.ac.uk). This equipment was funded by BIS National E-infrastructure capital grant ST/K00042X/1, STFC capital grants ST/H008519/1 and ST/K00087X/1, STFC DiRAC Operations grant ST/K003267/1 and Durham University. DiRAC is part of the National E-Infrastructure.
Numerical computations presented in this paper were in part carried out on the general-purpose PC farm at Center for Computational Astrophysics, CfCA, of National Astronomical Observatory of Japan.





\begin{thebibliography}{99}
\bibitem[\protect\citeauthoryear{Planck Collaboration}{2016}]{Planck15} Ade P. A. R., Aghanim N., Arnaud M., Ashdown M., Aumont J., Baccigalupi C. et al., 2016, A~\&~A, 594, 13
\bibitem[\protect\citeauthoryear{Alonso}{2012}]{cute}
Alonso D., 2012, arXiv:1210.1833 [astro-ph.IM]
\bibitem[\protect\citeauthoryear{Arnold~et~al.}{2014}]{arnold14}
Arnold C., Puchwein E., Springel V., 2014, MNRAS, 440, 833
\bibitem[\protect\citeauthoryear{Barreira et al.}{2017}]{barreira17} Barreira A., Bose S., Li B., Llinares C., 2017, JACP, 02, 031
\bibitem[\protect\citeauthoryear{Barreira et al.}{2015a}]{barreira15a} Barreira A., Cautun M., Li B., Baugh C.~M., Pascoli S., 2015a, JACP, 08, 028
\bibitem[\protect\citeauthoryear{Barreira et al.}{2015b}]{barreira15b} Barreira A., Li B., Jennings E., Merten  J., King L., Baugh C. M., Pascoli S., 2015b, MNRAS, 454, 4085
\bibitem[\protect\citeauthoryear{Bartelman \& Schneider}{2001}]{bartelman01}
Bartelman M., Schneider P., 2001, Phys. Rept., 340, 291
\bibitem[\protect\citeauthoryear{Behroozi et al.}{2013}]{rockstar}
Behroozi P.~S., Wechsler R.~H., Wu H., 2013, ApJ, 762, 109
\bibitem[\protect\citeauthoryear{Berlind et al.}{2003}]{hod}
Berlind A.~A., Weinberg D.~H., Benson A.~J., Baugh C.~M., Cole S., Dave R., Frenk C.~S., Jenkins A., Katz N., Lacey C.~G., 2003, ApJ, 593, 1
\bibitem[\protect\citeauthoryear{Bose et al.}{2015}]{bose15}
Bose S., Hellwing W.~A., Li B., 2015, JCAP, 02, 034
\bibitem[\protect\citeauthoryear{Bose et al.}{2017}]{bose17}
Bose S., Li B., Barreira A., He J., Hellwing W.~A., Koyama K., Llinares C., Zhao G., 2017, JCAP, 02, 050
\bibitem[\protect\citeauthoryear{Brainerd et al.}{1996}]{1996ApJ...466..623B}
Brainerd T. G., Blandford R. D., Smail I., 1996, ApJ, 466, 623
\bibitem[\protect\citeauthoryear{Brax et al.}{2008}]{Brax2008}
Brax P., van de Bruck C., Davis A.-C., Shaw D. J., 2008, PRD, 78, 104021
\bibitem[\protect\citeauthoryear{Cai~et~al.}{2015}]{cai15}
Cai Y., Padilla N., Li B., 2015, MNRAS, 451, 1036
\bibitem[\protect\citeauthoryear{Carroll~et~al.}{2004}]{carroll1}
Carroll S.~M.; Duvvuri V. Trodden M., Turner M.~S., 2004, PRD, 70, 043528
\bibitem[\protect\citeauthoryear{Carroll~et~al.}{2005}]{carroll2}
Carroll S.~M., de Felice A., Duvvuri V., Easson D.~A., Trodden M., Turner M.~S., 2005, PRD, 71, 063513
\bibitem[\protect\citeauthoryear{Cataneo~et~al.}{2015}]{cataneo15}
Cataneo M., Rapetti D., Schmidt F., Mantz A.~B., Allen S.~W., Applegate D.~E., Kelly P.~L., von der Linden A., Morris R.~G., 2015, PRD, 92, 044009
\bibitem[\protect\citeauthoryear{Cautun~et~al.}{2017}]{cautun17}
Cautun M., Paillas E., Cai Y., Bose S., Armijo J., Li B., Padilla N., 2017, MNRAS, submitted; arXiv:1710.01730 [astro-ph.CO]
\bibitem[\protect\citeauthoryear{Ceron-Hurtado et al.}{2016}]{Ceron-Hurtado-2016}
Ceron-Hurtado J. J., He J., Li B., 2016, PRD, 94, 064052
\bibitem[\protect\citeauthoryear{Clampitt et al.}{2011}]{clampitt17}
Clampitt J., Sanchez C., Kwan J., Krause E., MacCran N., Park Y., et al., 2017, MNRAS, 465, 4204 
\bibitem[\protect\citeauthoryear{Colombi \& Novikov}{2011}]{powmes}
Colombi S., Novikov D., 2011, Astrophysics Source Code Library, record ascl:1110.017
\bibitem[\protect\citeauthoryear{Crocce et al.}{2006}]{crocce06} Crocce M., Pueblas S., Scoccimarro R., 2006, MNRAS, 373, 369
\bibitem[\protect\citeauthoryear{Deffayet et al.}{2009}]{deffayet09} Deffayet C., Esposito-Farese G., Vikman A., PRD, 79, 084003
\bibitem[\protect\citeauthoryear{DES Collaboration}{2017}]{DESY1}
DES Collaboration et al., 2017, preprint (arXiv:1708.01530)
\bibitem[\protect\citeauthoryear{Dirian et al.}{2014}]{dirian14} Dirian Y., Foffa S., Khosravi N., Kunz M., Maggiore M., JCAP, 06, 033
\bibitem[\protect\citeauthoryear{Fang et al.}{2017}]{fang17}
Fang W., Li B., Zhao G., 2017, PRL, 118, 181301
\bibitem[\protect\citeauthoryear{Glayzes et al.}{2015}]{gleyzes15} Gleyzes J., Langlois D., Piazza F., Vernizzi F., PRL, 114, 211101
\bibitem[\protect\citeauthoryear{Guzik \& Seljak}{2002}]{2002MNRAS.335..311G}
Guzik J., Seljak U., 2002, MNRAS, 335, 311
\bibitem[\protect\citeauthoryear{Harnois-Deraps et al.}{2015}]{Harnois15}
Harnois-Deraps J., Munshi D., Valageas P., van Waerbeke L., Brax P., Coles P., Rizzo L., 2015, MNRAS, 454, 2722
\bibitem[\protect\citeauthoryear{He et al.}{2016}]{he16}
He J., Li B., Baugh C.~M., 2016, PRL, 117, 221101
\bibitem[\protect\citeauthoryear{Higuchi \& Shirasaki}{2016}]{higuchi16}
Higuchi Y., Shirasaki M., 2016, MNRAS, 459, 2762
\bibitem[\protect\citeauthoryear{Hilbert et al.}{2009}]{hilbert08}
Hilbert S., Hartlap J., White S.~D.~M., Schneider P., 2009, A{\&}A, 499, 31
\bibitem[\protect\citeauthoryear{Hildebrandt et al.}{2017}]{kids450}
H. Hildebrandt et al., 2017, MNRAS, 465, 1454
\bibitem[\protect\citeauthoryear{WMAP Collaboration}{2013}]{WMAP9} Hinshaw G., Larson D., Komatsu E., Spergel D.~N., Bennett C.~L., Dunkley J., Nolta M.~R., Halpern M., Hill R.~S., Odegard N., Page L., Smith K.~M., Weiland J.~L., Gold B., Jarosik N., Kogut A., Limon M., Meyer S.~S., Tucker G.~S., Wollack E., Wright E.~L., 2013, ApJS, 208, 19
\bibitem[\protect\citeauthoryear{Hirata \& Seljak}{2004}]{hirata04}
Hirata C. M., Seljak U., 2004, PRD, 70, 063526
\bibitem[\protect\citeauthoryear{Hoekstra et al.}{2004}]{2004ApJ...606...67H}
Hoekstra H., Yee H. K. C., Gladders M. D., 2004, ApJ, 606, 67
\bibitem[\protect\citeauthoryear{Hoekstra \& Jain}{2008}]{hoekstra08} Hoekstra H., Jain B., Ann. Rev. Nucl. Part. Sci., 58, 99
\bibitem[\protect\citeauthoryear{Hu \& Sawicki}{2007}]{hs2007} Hu W., Sawicki I., 2007, PRD, 76, 064004
\bibitem[\protect\citeauthoryear{Hudson et al.}{1998}]{1998ApJ...503..531H}
Hudson M. J., Gwyn S. D. J., Dahle H., Kaiser N., 1998, ApJ, 503, 531
\bibitem[\protect\citeauthoryear{Huterer et al.}{2006}]{Huterer06}
D. Huterer, M. Takada, G. Bernstein, and B. Jain, 2006, MNRAS, 366, 101,
\bibitem[\protect\citeauthoryear{Jain et al.}{2000}]{jsw}
Jain B., Seljak U., White S.~M.~D., 2000, ApJ, 530, 547
\bibitem[\protect\citeauthoryear{Jeong et al.}{2009}]{jeong09}
Jeong D., Komatsu E., Jain B., 2009, PRD, 80, 123527
\bibitem[\protect\citeauthoryear{Joyce et al.}{2015}]{joyce15} Joyce A., Jain B., Khoury J., Trodden M., 2015, Phys. Rept., 568, 1
\bibitem[\protect\citeauthoryear{Khoury \& Weltman}{2004}]{chameleon}
Khoury J., Weltman A., 2004, PRD, 69, 044026 
\bibitem[\protect\citeauthoryear{Kilbinger}{2015}]{kilbinger15} 
Kilbinger M., 2015, Rep. Prog. Phys., 78, 086901
\bibitem[\protect\citeauthoryear{Koyama}{2016}]{Koyama2016} 
Koyama K., 2016, Rep. Prog. Phys., 79, 046920
\bibitem[\protect\citeauthoryear{Li et al.}{2013}]{Li_PK_2013}
Li B., Hellwing W.~A., Koyama K., Zhao G., Jennings E., Baugh C.~M., 2013, MNRAS, 428, 743
\bibitem[\protect\citeauthoryear{Li et al.}{2012}]{ecosmog}
Li B., Zhao G., Teyssier R., Koyama K., 2012, JCAP, 01, 051
\bibitem[\protect\citeauthoryear{Liu et al.}{2016}]{liu16}
Liu X., Li B., Zhao G., Chiu M., Fang W., Pan C., Wang Q., Du W., Yuan W., Fu L., Fan Z., PRL, 117, 051101
\bibitem[\protect\citeauthoryear{Lewis \& Challinor}{2011}]{camb} Lewis A., Challinor A., 2011, Astrophysics Source Code Library, record ascl:1102.026
\bibitem[\protect\citeauthoryear{Maggiore \& Mancarella}{2014}]{maggiore14} Maggiore M., Mancarella M., PRD, 90, 023005
\bibitem[\protect\citeauthoryear{Mandelbaum et al.}{2006a}]{2006MNRAS.368..715M}
Mandelbaum R., Seljak U., Kauffmann G., Hirata C. M., Brinkmann J., 2006, MNRAS, 368, 715
\bibitem[\protect\citeauthoryear{Mandelbaum et al.}{2006b}]{mandelbaum06}
Mandelbaum R., Seljak U., Cool R., Blanton M., Hirata C.~M., Brinkmann J., 2006, MNRAS, 372, 758 
\bibitem[\protect\citeauthoryear{Mandelbaum et al.}{2008}]{mandelbaum08}
Mandelbaum R., Seljak U., Hirata C.~M., 2008, JCAP, 08, 06
\bibitem[\protect\citeauthoryear{Manera et al.}{2015}]{manera2015}
Manera M., Samushia L., Tojeiro R., Howlett C., Ross A.~J., Percival W.~J., Gil-Marin H., Brownstein J.~R., Burden A., Montesano F., 2017, MNRAS, 447, 437
\bibitem[\protect\citeauthoryear{Mota \& Shaw}{2007}]{chameleon2}
Mota D.~F., Shaw D.~J., 2007, PRD, 75, 063501
\bibitem[\protect\citeauthoryear{Mummery~et~al.}{2017}]{mummery17}
Mummery B., McCarthy I.~G., Bird S., Schaye J., 2017, arXiv:1702.02064 [astro-ph.CO]
\bibitem[\protect\citeauthoryear{Navarro~et~al.}{1997}]{NFW}
Navarro J., Frenk C.~S., White S.~D.~M., 1997, ApJ, 490, 493
\bibitem[\protect\citeauthoryear{Nicolis et al.}{2009}]{nicolis09} Nicolis A., Rattazzi R., Trincherini E., PRD, 79, 064036
\bibitem[\protect\citeauthoryear{Oguri \& Takada}{2011}]{oguri11}
Oguri M., Takada M., 2011, PRD, 83, 023008
\bibitem[\protect\citeauthoryear{Osato~et~al.}{2015}]{osato15}
Osato K., Shirasaki M., Yaoshida N., 2015, ApJ, 806, 186
\bibitem[\protect\citeauthoryear{Peirone~et~al.}{2017}]{peirone17}
Peirone S., Raveri M., Viel M., Borgani S., Ansoldi S., 2017, PRD, 95, 023521
\bibitem[\protect\citeauthoryear{Perlmutter~et~al.}{1999}]{perlmutter99}
Perlmutter S., Aldering G., Goldhaber G., Knop R.~A., Nugent P., Castro P.~G., et al., 1999, ApJ, 517, 565
\bibitem[\protect\citeauthoryear{Pratten~et~al.}{2016}]{Pratten16}
Pratten G., Munshi D., Valageas P., Brax P., 2016, PRD, 93, 103524
\bibitem[\protect\citeauthoryear{Refregier}{2003}]{refregier03}
Refregier A., 2003, Ann. Rev. Astron. Astrophys., 41, 645
\bibitem[\protect\citeauthoryear{Riess et al.}{1998}]{riess98} Riess A.~G., Filippenko A.~V., Challis P., Clocchiatti A., Diercks A., Garnavich P.~M., et al., 1998, ApJ, 116, 1009
\bibitem[\protect\citeauthoryear{Sato et al.}{2009}]{Sato09}
Sato M., Hamana T., Takahashi R., Takada M., Yoshida, N., Matsubara, T.,
Sugiyama N., 2009, ApJ, 701, 945
\bibitem[\protect\citeauthoryear{Semboloni~et~al.}{2011}]{semboloni11}
Semboloni E., Hoekstra H., Schaye J., van Daalen M.~P., McCarthy I.~G., 2011, MNRAS, 417, 2020
\bibitem[\protect\citeauthoryear{Sheldon~et~al.}{2004}]{sheldon04}
Sheldon E.~S., Johnston D.~E., Frieman J.~A., Scranton R., McKay T.~A., Connolly A.~J., Budavari T., Zehavi I., Bahcall N.~A., Brinkmaann J., Fukugita M., 2004, ApJ, 127, 2544
\bibitem[\protect\citeauthoryear{Shi~et~al.}{2017}]{shi17}
Shi D., Li B., Han J., 2017, MNRAS, in press
\bibitem[\protect\citeauthoryear{Shi~et~al.}{2015}]{shi15}
Shi D., Li B., Han J., Gao L., Hellwing W.~A., 2015, MNRAS, 452, 3179
\bibitem[\protect\citeauthoryear{Shirasaki~et~al.}{2015}]{shirasaki15}
Shirasaki M., Hamana T., Yoshida N., 2015, MNRAS, 454, 3043
\bibitem[\protect\citeauthoryear{Shirasaki~et~al.}{2016}]{shirasaki16}
Shirasaki M., Hamana T., Yoshida N., 2016, PASJ, 68, 4
\bibitem[\protect\citeauthoryear{Shirasaki~et~al.}{2017}]{shirasaki17}
Shirasaki M., Nishimichi T., Li B., Higuchi Y., 2017, MNRAS, 466, 2402
\bibitem[\protect\citeauthoryear{Smith~et~al.}{2003}]{halofit}
Smith R.~E., Peacock J., Jenkins A., White S.~D.~M., Frenk C.~S., Pearce F.~R., Thomas P.~A., Efstathiou G., Couchman, H.~M.~P., 2003, MNRAS, 341, 1311
\bibitem[\protect\citeauthoryear{Takahashi~et~al.}{2012}]{halofit2}
Takakashi R., Sato M., Nishimichi T., Taruya A., Oguri M., 2012, ApJ, 761, 152
\bibitem[\protect\citeauthoryear{Tessore~et~al.}{2015}]{tessore15}
Tessore N., Winther H.~A., Metcalf R.~B., Ferreira P.~G., Giocoli C., 2015, JCAP, 10, 036
\bibitem[\protect\citeauthoryear{Teyssier}{2002}]{ramses}
Teyssier R., 2002, A~\&~A, 385, 337
\bibitem[\protect\citeauthoryear{Wang et al.}{2012}]{Wang2012}
Wang J., Hui L., Khoury J., 2012, PRL, 109, 241301
\bibitem[\protect\citeauthoryear{Yoo et al.}{2006}]{yoo2006}
Yoo J., Tinker J.~L., Weinberg D.~H., Zheng Z., Katz N., Dave R., 2006, ApJ, 652, 26
\bibitem[\protect\citeauthoryear{Zheng et al.}{2005}]{zheng2005}
Zheng Z., Berlind A.~A., Weinberg D.~H., Benson A.~J., Baugh C.~M., Cole S., Dave R., Frenk C.~S., Jenkins A., Katz N., Lacey C.~G., 2005, ApJ, 633, 791
\end{thebibliography}







\bsp	
\label{lastpage}
\end{document}